\documentclass[journal=jctcce,manuscript=article,articletitle=true]{achemso}
\usepackage[version=4]{mhchem}
\usepackage{comment}
\usepackage{float}
\usepackage{placeins} 
\usepackage{xcolor}
\usepackage{soul}
\usepackage{multirow}
\usepackage{makecell}
\usepackage{threeparttable}
\usepackage{tabularx}

\author{Natascia L. Fragapane}
\affiliation{Inorganic Chemistry Laboratory, Department of Chemistry, University of Oxford,\\ Oxford OX1 3QR, United Kingdom}
\author{Volker L. Deringer}
\affiliation{Inorganic Chemistry Laboratory, Department of Chemistry, University of Oxford,\\ Oxford OX1 3QR, United Kingdom}
\email{volker.deringer@chem.ox.ac.uk}

\title{Li--P--S Electrolyte Materials as a Benchmark for Machine-Learned Interatomic Potentials}

\begin{document}

\begin{abstract}
  With the growing availability of machine-learned interatomic potential (MLIP) models for materials simulations, there is an increasing demand for robust, automated, and chemically insightful benchmarking methodologies.
  In response, we here introduce LiPS-25, a curated benchmark dataset for a canonical series of solid-state electrolyte materials from the \ce{Li2S}--\ce{P2S5} pseudo-binary compositional line, including crystalline and amorphous configurations. Together with the dataset, we present a suite of performance tests that range from conventional numerical error metrics to physically motivated evaluation tasks. 
  With a focus on graph-based MLIP architectures, we run numerical experiments that assess (i)~the effect of hyperparameters and (ii)~the fine-tuning behavior of selected pre-trained (``foundational'') MLIP models.
  Beyond the Li--P--S solid-state electrolytes, we expect that such benchmarks and their code implementations can be readily adapted to other material systems.
  
\end{abstract}

\setcounter{section}{0}
\section*{Introduction}

Machine-learned interatomic potentials (MLIPs) are now a standard tool for atomistic simulation, offering first-principles accuracy at a fraction of the computational cost \cite{deringer_machine_2019, friederich_machine-learned_2021, behler_four_2021, unke_machine_2021}. They have enabled a new degree of realism in materials modeling, with applications including device-scale simulations \cite{zhou_device-scale_2023}, long-timescale dynamics \cite{charron_navigating_2025}, and high-throughput materials screening \cite{merchant_scaling_2023}. More recently, pre-trained or ``foundation'' models \cite{chen_universal_2022, takamoto_towards_2022, deng_chgnet_2023, batatia_foundation_2025, yang_mattersim_2024, park_scalable_2024, zhang_dpa-2_2024, zhang_pretraining_2024, neumann_orb_2024, rhodes_orb-v3_2025} have further lowered the barrier to entry: trained on large and diverse datasets, they can be applied to new systems with little to no additional training, dramatically reducing the time and expertise required as compared to crafting MLIPs by hand. 

As fitting architectures and specific models continue to proliferate, the systematic and automated benchmarking of MLIPs is becoming ever more important: for identifying state-of-the-art models, clarifying their strengths and limitations, and setting standards for reproducibility and comparison across the field. Several datasets have become widely adopted for evaluating MLIPs \cite{jurecka_benchmark_2006, rezac_s66_2011, rezac_extensions_2011, ruddigkeit_enumeration_2012, ramakrishnan_quantum_2014, chmiela_machine_2017, smith_ani-1_2017, wu_moleculenet_2018, smith_ani-1ccx_2020, zuo_performance_2020, chanussot_open_2021, tran_open_2023, pengmei_beyond_2024, pota_thermal_2024}. Benchmarks such as QM9 \cite{ramakrishnan_quantum_2014}, MD17 \cite{chmiela_machine_2017, pengmei_beyond_2024}, and MoleculeNet \cite{wu_moleculenet_2018} have provided insight into model performance on static molecular properties, typically evaluated using established metrics such as the root-mean-square error (RMSE) or mean absolute error (MAE). These metrics, albeit a necessary first step, are not always indicative of accurate model performance in downstream simulations \cite{morrow_indirect_2022, fu_forces_2023, liu_discrepancies_2023}. More physically motivated benchmark tasks have since begun to emerge, for instance in datasets such as OC20/22 \cite{chanussot_open_2021, tran_open_2023}, frameworks such as MLIPX \cite{zills_mlipx_2025}, and leaderboard platforms such as Matbench \cite{dunn_benchmarking_2020, riebesell_framework_2025} or JARVIS \cite{choudhary_jarvis-leaderboard_2024}. And still, comprehensively assessing the robustness and transferability of MLIP models for real-world modeling applications remains a challenge. \cite{omee_structure-based_2024, tawfik_computational_2025}. 

One such application is the atomistic modeling of lithium thiophosphates (``LiPS'' in the following). The LiPS family is a prototypical solid-state electrolyte (SSE) system, combining high ionic conductivity \cite{kamaya_lithium_2011, seino_sulphide_2014, kato_high-power_2016, kraft_inducing_2018,
janek_challenges_2023}, a wide electrochemical stability range, and low cost \cite{famprikis_fundamentals_2019}. Materials along the \ce{Li2S}--\ce{P2S5} compositional line (Figure \ref{fig:lips-25}a) are of particular interest due to the variety of phases that are accessible depending on preparation conditions: from crystalline to glassy--ceramic and fully amorphous. Li--P--S phases have been the subject of extensive experimental \cite{eckert_structural_1990, hayashi_preparation_2004, mizuno_new_2005, mizuno_high_2006, sakuda_sulfide_2013, ohara_structural_2016, dietrich_lithium_2017, tsukasaki_direct_2017, kudu_review_2018, chen_sulfide_2018, garcia-mendez_correlating_2020} and computational \cite{sistla_structural_2004, onodera_reverse_2011, mori_visualization_2013, zhu_origin_2015, baba_structure_2016, chang_super-ionic_2018, kim_atomistic_2019, smith_low-temperature_2020, sadowski_computational_2020, ohkubo_conduction_2020, hajibabaei_universal_2021, forrester_disentangling_2022, ariga_new_2022, guo_artificial_2022, staacke_tackling_2022, gigli_mechanism_2024} investigation. Their structural complexity makes them a suitable test system for MLIPs: a successful model must capture diverse atomic environments and complex dynamic properties arising from those \cite{xu_progress_2022}. 

Indeed, several recent works have begun exploring benchmarking approaches tailored to Li-ion conductors and SSEs. Therrien et al.\ introduced a curated dataset of SSE materials with experimental ionic conductivities, applying it to assess the performance of various MLIPs \cite{therrien_obelix_2025}. Dembitskiy et al.\ developed LiTraj, a dataset focused on Li-ion migration barriers, which enabled comparison of different ML models on property-prediction tasks and demonstrated the impact of fine-tuning for foundation models \cite{dembitskiy_benchmarking_2025}. A recent framework introduced by Du et al.\ broadens the scope of assessment, incorporating properties such as the bulk modulus, energy above the convex hull, and Li-ion diffusion, enabling a systematic evaluation of pre-trained models \cite{du_assessment_2025}. Together, these studies mark important progress in creating physically grounded and application-relevant frameworks for assessing MLIPs for SSEs. Yet, a key gap remains: first-principles-labeled benchmarking datasets that enable end-to-end assessment of MLIPs across a structurally diverse set of configurations. Such datasets, consisting of atomistic structures labeled with energies and forces rather than only property-level data, would support comprehensive evaluation of MLIPs on both static and dynamic properties and enable systematic studies of fine-tuning -- a strategy shown to be critical for accurately capturing structure and dynamics in glassy SSEs \cite{bertani_atomic_2025}.

Here, we present LiPS-25, a curated dataset of crystalline and amorphous structures, designed to support rigorous evaluation of MLIPs. We envisage its utility to be two-fold: as a general-purpose, off-the-shelf dataset for modeling materials across the \ce{Li2S}--\ce{P2S5} tie-line; and, beyond this, as an application-relevant benchmarking tool for ML potentials. To this end, we present a suite of accompanying performance tests that extend beyond conventional error metrics to physically motivated evaluations. Together with these tests, LiPS-25 provides a benchmark platform for both MLIP research and its practical applications.

\FloatBarrier

\section*{The LiPS-25 Dataset}

\begin{figure*}[htbp]
    \centering
    \includegraphics[width=\linewidth]{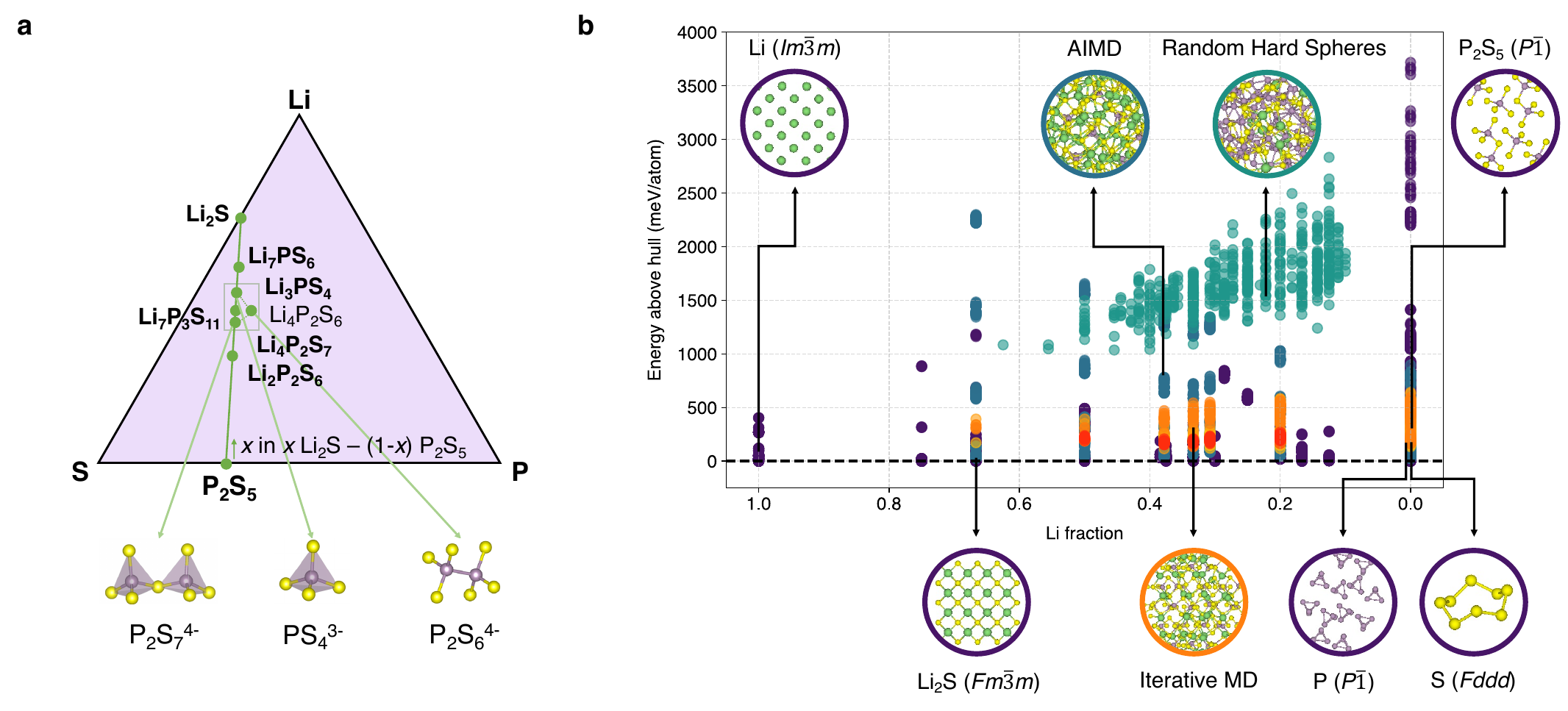}
    \caption{The LiPS-25 dataset. 
    (a) Ternary diagram of the Li--P--S system, with the tie-line between \ce{Li2S} and \ce{P2S5} indicated. Green circles mark compositions with known crystalline phases; compositions in bold were used to build the LiPS-25 dataset. Key structural motifs are displayed below: \textit{ortho}-thiophosphate, [\ce{PS4}]$^{3-}$; \textit{pyro}-thiophosphate, [\ce{P2S7}]$^{4-}$; and \textit{hypo}-thiophosphate, [\ce{P2S6}]$^{4-}$.
    (b) Scatter plot for the LiPS-25 dataset showing the fraction of Li in each structure ($x$-axis) versus energy above the convex hull ($y$-axis); a dashed line at $y=0$ has been added. Dimer configurations are excluded from this plot. Representative structures are shown (atomic color coding: Li, green; P, purple; S, yellow), visualized with VESTA \cite{momma_vesta_2011}; colored outlines act as a legend for the scatter plot, with purple for crystalline structures, teal for AIMD snapshots, orange/red for iterative melt–quench configurations (Iter1-$x$, Iter2-$x$), and turquoise for random hard spheres.
    }
    \label{fig:lips-25}
\end{figure*}

The LiPS-25 dataset was curated relying on domain knowledge to cover relevant compositions and polymorphs along the pseudo-binary \ce{Li2S}--\ce{P2S5} tie-line, focusing on 7 key compositions (Figure \ref{fig:lips-25}) atop a broader coverage of the Li--P--S phase space. DFT energy and force reference data (``labels'') are calculated at the PBEsol level \cite{perdew_restoring_2008}, the latter chosen based on previous benchmarking studies for \ce{Li3PS4} (ref.~\citenum{gigli_mechanism_2024}) and related \ce{Li10GeP2S12}-type ion conductors \cite{huang_deep_2021}. In ref.~\citenum{huang_deep_2021}, PBEsol was shown to accurately reproduce experimental lattice parameters, and the predicted lithium diffusion coefficients were found to be largely insensitive to the choice between PBE and PBEsol. In ref.~\citenum{gigli_mechanism_2024}, PBEsol was found to perform comparably with more computationally expensive alternatives, namely, the meta-GGA r$^2$SCAN and the hybrid PBE0 exchange--correlation functionals, in predicting dynamic and phase transition behaviors. While PBE0 more faithfully reproduced electronic band gaps, this is less relevant for the present study, which focuses on atomistic structural and dynamic properties rather than the electronic structure. 

The initial dataset before iterative training started (which we call ``Iter0'') was designed to provide a sufficiently robust starting point for modeling a diverse range of atomic environments, enabling subsequent targeted iterations of data collection. The Iter0 dataset consists of several components: distorted and ``rattled'' elemental, binary, and ternary Li/P/S crystalline structures taken from the Inorganic Crystal Structure Database (ICSD) \cite{zagorac_recent_2019} and the Materials Project (MP) \cite{jain_commentary_2013, horton_accelerated_2025}; snapshots from ab initio molecular dynamics (AIMD) simulations at 250, 500, and 1000 K for each of the 7 key compositions considered along the tie-line (\ce{Li2S}, \ce{Li7PS6}, \ce{Li3PS4}, \ce{Li7P3S11}, \ce{Li4P2S7}, \ce{Li2P2S6}, and \ce{P2S5}; Figure \ref{fig:lips-25}a); random-hard-sphere structures generated using \texttt{buildcell} \cite{pickard_ab_2011}; and isolated dimer configurations of every combination of Li, P, and S atoms. 

Iterative melt--quench (MQ) simulations, using the NequIP architecture, \cite{batzner_e3-equivariant_2022} were then used to extend the dataset. The objective of these iterations is to augment Iter0 by including liquid and amorphous phases, as well as to extend the sampling of disordered crystalline structures. Iterative MQ simulations were carried out with a query-by-committee procedure to select the most ``uncertain'', and thus most informative, structures at each iteration. The MQ protocols were applied to the 7 key compositions and span two main iterations. Iter1-(1–4) forms a more general addition to the dataset, with NVT MQ cycles of 300~K $\rightarrow T_{\text{melt}} \rightarrow 300$~K ($T_{\text{melt}}$ = 1000, 1500 K; quench rates 50, 100 K/ps); the most uncertain structures across all trajectories were added to the dataset. Iter2-(1–3) instead focused on augmenting the dataset with glassy structures only. From NPT MQ cycles of 300~K $\rightarrow$ 1500~K $\rightarrow T_{\text{quench}}$  ($T_{\text{quench}}$ = 300, 400, 500 K; quench rate 50 K/ps), data selection was restricted to the most uncertain structures only within the anneal post melt--quench, thus specifically targeting amorphous structures.

\begin{table}[t]
    \caption{Composition of the LiPS-25 dataset. Columns report the number of cells, $N_\text{cells}$, and the total number of atoms, $N_\text{atoms}$, of each structure type, along with the average energy above the convex hull, $\overline{E}_{\mathrm{hull}}$, and the 10th--90th percentile range of energies above the hull, $P_{90-10}(E_{\mathrm{hull}})$.}
    \centering
    \begin{threeparttable}
        \begin{tabular}{lcccc}
            \hline
            \textbf{Data type} & 
            \textbf{\textit{N}$_{\!\textbf{cells}}$} & 
            \textbf{\textit{N}$_{\!\textbf{atoms}}$} & 
            \textbf{\boldmath$\overline{E}_{\mathrm{hull}}$} & 
            \textbf{\boldmath$P_{90-10}(E_{\mathrm{hull}})$} \\
            & & & \textbf{(meV/atom)} & \textbf{(meV/atom)} \\
            \hline
            Crystalline   & 8,891 & 258,880 & 176 & 432 \\ 
            AIMD          & 1,246 & 103,301 & 638 & 1,300 \\ 
            Random Hard Spheres & 500 & 50,553 & 1,673 & 673 \\ 
            Dimers        & 138   & 276     & 4,840 & 5,225 \\ 
            \hline
            Iter1\tnote{1} & 1,000 & 52,217 & 336 & 325 \\ 
            Iter2\tnote{1} & 750   & 66,349 & 205 & 74 \\ 
            \hline
            \textbf{Total} & 12,525 & 531,576  & 348 & 1,007 \\ 
            \hline
        \end{tabular}
        \begin{tablenotes}
            \footnotesize
            \item[1] ``Iter1'' and ``Iter2'', respectively, refer to all Iter1-\textit{x} and Iter2-\textit{x} datasets combined.
        \end{tablenotes}
    \end{threeparttable}
    \label{tab:dataset_breakdown}
\end{table}

The components of the LiPS-25 dataset are visualized in Figure \ref{fig:lips-25}b and summarized in Table \ref{tab:dataset_breakdown}, and further details are provided in the Supporting Information.
LiPS-25 includes pre-defined training, validation, and test sets for cross-comparability and consistent model evaluation. The validation set is used to tune hyperparameters and assess model performance during training, whereas the test set is used to evaluate model performance after the training is complete. To ensure that each subset represents the diversity of the complete dataset, we employ random stratified sampling with an 80:10:10 split. The dataset is openly available \cite{fragapane_lips-25_2025}.

\section*{Benchmark Tasks}

\begin{figure*}[htbp]
    \centering
    \includegraphics[width=\linewidth]{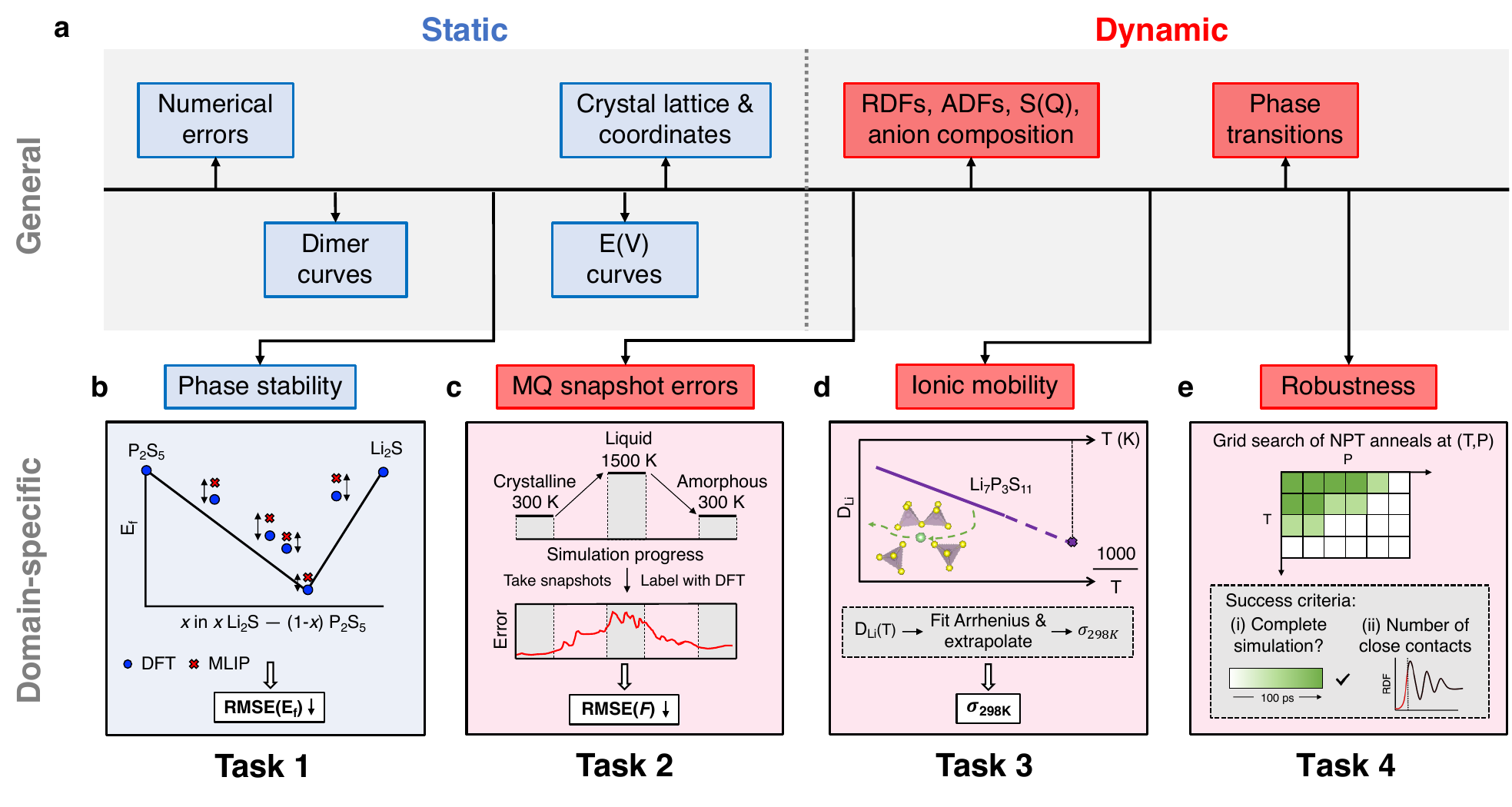}
    \caption{Benchmark tasks. 
    (a) Overview of validation techniques for MLIPs, which fall into two groups: ``static'' validation, assessing numerical errors and basic energetic profiles, and ``dynamic'' validation based on MD simulations, leveraging domain expertise to evaluate MLIP performance. From these, four domain-specific benchmarking tasks have been selected to accompany the LiPS-25 dataset, facilitating physically motivated evaluation of MLIPs. (b) \textbf{Task 1:} Energetic accuracy. The MLIP is used to predict formation energies of 8 crystalline structures along the \ce{Li2S}--\ce{P2S5} tie-line. These predictions are compared against ground-truth values to compute the RMSE(E$_{f}$) metric. (c) \textbf{Task 2:} Domain-specific force accuracy. Evenly-spaced snapshots are taken from an NPT melt--quench simulation of \ce{Li7P3S11}. Force errors are calculated with respect to DFT, and aggregated into the RMSE(\textit{F}) metric. (d) \textbf{Task 3:} Property accuracy. The room-temperature ionic conductivity of the \ce{Li7P3S11} crystal, a known superionic conductor, is predicted; the inset illustrates Li ion migration, visualized with VESTA \cite{momma_vesta_2011}. (e) \textbf{Task 4:} Robustness. NPT simulations across a grid of temperatures and pressures are run and assessed by simulation survival and by the number of close-contact events.
    }
    \label{fig:figure_2}
\end{figure*}

To accompany the LiPS-25 dataset, we introduce a set of four benchmark tasks (Figure \ref{fig:figure_2}) that evaluate MLIP performance on key aspects relevant to SSE modeling. These physically motivated evaluations provide broad yet informative indicators of model suitability, including physical accuracy and dynamic fidelity, that complement static errors. By extending beyond conventional numerical metrics, the benchmarks are able to capture subtle limitations that can affect a model’s applicability to specific materials systems and applications. Each task is designed to balance computational cost with diagnostic value, enabling the systematic comparison of multiple models without the (often prohibitive) computational expense of full-scale production simulations that might demand longer timescales or larger simulation sizes. Crucially, we focus our tasks on physically meaningful observables for which reliable experimental or computational reference data already exist or can be reasonably collected. This focus ensures that the resulting comparisons are both grounded and interpretable within the broader context of SSE research.

For each task, we describe the aspect of MLIP performance being assessed -- such as accuracy on the LiPS-25 dataset, generalizability to out-of-domain configurations, or robustness -- along with the  methodology used and the relevance of the task to SSE modeling.
All benchmarks are accompanied by Python notebooks, and, where relevant, LAMMPS input files, to ensure reproducibility and to facilitate the application of these benchmarks to other MLIPs.

\subsection*{Task 1: Energetic accuracy}
This task assesses the accuracy with which an MLIP reproduces DFT formation energies. In contrast to standard energy MAE/RMSE performance metrics, this task requires both reliable force predictions to obtain correct relaxed geometries and accurate energies to evaluate stability. Formation energies provide a chemically informative measure of relative phase stability (SSEs must be thermodynamically stable under operating conditions to avoid undesired decomposition or phase transitions that would degrade their performance), synthesizability (knowledge of the relative formation energies of competing phases can help identify compositions that are more likely to be experimentally feasible), and reactivity (e.g., electrolyte decomposition at electrode interfaces), all of which directly impact electrolyte performance. 

The formation energy per atom, $E_{\text{f/atom}}$, is calculated relative to the end-points, \ce{Li2S} and \ce{P2S5}, for selected compositions along \textit{x}\ce{Li2S}--(1--\textit{x})\ce{P2S5}, as in ref.~\citenum{guo_artificial_2022}:
\begin{equation}
    E_{\text{f/atom}} = \frac{1}{7 - 4x} \left[ E_{(\ce{Li2S})_x(\ce{P2S5})_{1-x}} - x  E_{\ce{Li2S}} - (1 - x) E_{\ce{P2S5}} \right].
    \label{eq:formation_energy}
\end{equation}

$E_{\text{f/atom}}$ values are obtained from the crystal structures relaxed with the corresponding method (MLIP or DFT), and the RMSE($E_{\text{f}}$) is computed by comparing the MLIP and DFT energy labels. In this method-specific relaxation approach, the RMSE($E_{\text{f}}$) conflates energetic and force accuracy, since models with poor force fidelity may converge to different geometries. To isolate single-point energetic accuracy, we also compute the RMSE($E_{\text{f}}$) from a common set of DFT-relaxed structures. These results are provided in the Supporting Information, and both evaluation protocols are implemented in a Jupyter notebook accompanying the present work.

The eight structures included in this task are \ce{Li2P2S6}, \ce{Li4P2S7}, \ce{Li7P3S11}, $\alpha$-\ce{Li3PS4}, $\beta$-\ce{Li3PS4}, $\gamma$-\ce{Li3PS4}, low-temperature \ce{Li7PS6} ($\mathit{Pna}2_1$)
, and high-temperature \ce{Li7PS6} ($\mathit{F}\overline{4}3\mathit{m}$). Of these structures, \ce{Li4P2S7}, $\alpha$-\ce{Li3PS4}, and high-temperature \ce{Li7PS6} have not been explicitly trained on, and thus present a test for an MLIP's extrapolation ability. 

\subsection*{Task 2: Domain-specific force accuracy}
This task evaluates the accuracy of an MLIP in predicting forces throughout a domain-specific molecular-dynamics (MD) simulation, as in prior work \cite{george_combining_2020, thomas_du_toit_hyperparameter_2024}. Here, DFT snapshots were computed every 5 ps from a NequIP-driven melt--quench trajectory of a 672-atom \ce{Li7P3S11} supercell between 300 and 1500 K. Force errors were then computed relative to these DFT labels. This test assesses the MLIP's (i) generalizability, through encompassing the full range of atomic environments relevant to the dataset -- fully ordered crystalline, through locally-ordered amorphous, to highly disordered liquid; and (ii) robustness, as the forces these snapshots experience are directly relevant to MD simulations -- thus poor prediction of these forces may indicate future unreliable propagation of dynamics. In contrast to Task~1, which evaluates models on optimized crystal structures only, Task~2 probes the model's accuracy in non-equilibrium, thermally disordered environments, thereby offering a more demanding and practically relevant evaluation of a given MLIP model.

\subsection*{Task 3: Property accuracy}
This task evaluates the ability of an MLIP to accurately capture lithium-ion diffusion, a property that emerges from accurate force predictions over extended timescales rather than being an explicit training target. Since lithium-ion mobility is central to SSE performance, models that fail to reproduce diffusion behavior would likely have limited practical value in realistic materials simulations, making this assessment a critical test of model applicability.

In this benchmark, we evaluate the ionic mobility in crystalline \ce{Li7P3S11}, a widely studied superionic conductor \cite{yamane_crystal_2007, minami_preparation_2010, seino_sulphide_2014}, which provides a rich basis for model validation. To minimize the computational expense of this task, we focus on a single representative composition and polymorph, with both the simulation length and box size converged (Figure S2). 500~ps NVT anneals across the relevant temperature regime of 400--800 K are carried out on 672-atom supercells. The diffusion coefficient at 298 K is extracted using the Arrhenius relation, and the corresponding ionic conductivity is estimated from the Nernst--Einstein relation. Full details of the calculation and discussion of this approach can be found in the Supporting Information.

\begin{table}[t]
\centering
\begin{threeparttable}
    \begin{tabular}{lllcc}
    \hline
        \textbf{Ref.} & \textbf{Method} & \textbf{Phase} & \textbf{$\sigma_{\text{RT}}$ (mS/cm)} & \textbf{$E_{\text{a}}$ (eV)} \\ \hline
        \citenum{seino_sulphide_2014} & Solid-state & Glass-ceramic & 0.08   & --   \\ 
        \citenum{seino_sulphide_2014} & Solid-state & Glass-ceramic & 1.4   & 0.50  \\ 
        \textbf{\citenum{seino_sulphide_2014}} & \textbf{Solid-state} & \textbf{Glass-ceramic}  & \textbf{17\tnote{a}}   & \textbf{0.17}  \\ 
        \citenum{chu_insights_2016} & Solid-state  & Glass-ceramic & 1.3   & 0.21 \\ 
        \citenum{chu_insights_2016} & Solid-state  & Glass-ceramic & 12   & 0.18 \\ \hline
        \citenum{mizuno_high_2006} & Mechanochemical & Glass-ceramic  & 3.2   & 0.12 \\ 
        \citenum{wenzel_interphase_2016} & Mechanochemical & Crystal  & 4   & 0.29 \\ 
        \citenum{busche_situ_2016} & Mechanochemical  & Crystal & 8.6   & 0.29 \\ \hline
        \citenum{ito_synthesis_2014} & Wet chemistry & Glass-ceramic & 0.27   & 0.39  \\ 
        \citenum{wang_mechanism_2018} & Wet chemistry & Glass-ceramic & 0.87   & 0.37 \\ 
        \citenum{calpa_preparation_2018} & Wet chemistry & Glass-ceramic & 0.011   & -- \\ 
        \citenum{calpa_preparation_2018} & Wet chemistry & Glass-ceramic & 1.0   & 0.13 \\ 
        \hline \hline
        \citenum{chu_insights_2016} & AIMD (PBE) & Crystal & 57.0  & 0.19 \\ 
        \textbf{\citenum{chu_insights_2016}} & \textbf{AIMD (PBEsol)} & \textbf{Crystal} & \textbf{61.0\tnote{b}}  & \textbf{--} \\ 
        \citenum{wang_computational_2017} & AIMD (PBE) & Crystal & 45.7 & 0.19 \\
        \citenum{chang_super-ionic_2018} & AIMD (PBE) & Crystal & 72.0  & 0.17 \\
        \citenum{sadowski_computational_2020} & AIMD (PBE) & Crystal & 84.0  & 0.17 \\ \hline
    \end{tabular}
    \begin{tablenotes}
        \footnotesize
        \item[a] Highest reported experimental conductivity \cite{seino_sulphide_2014}.
        \item[b] Representative AIMD value, computed using the same exchange--correlation functional as used for LiPS-25 \cite{chu_insights_2016}.
    \end{tablenotes}
\end{threeparttable}
\caption{Room-temperature ionic conductivities ($\sigma_{\text{RT}}$) and activation energies ($E_\text{a}$) for crystalline and glass-ceramic \ce{Li7P3S11} reported in the literature. Experimental values (top) are compiled based on the review in Ref.~\citenum{kudu_review_2018}, while computational values (bottom) were collected in this work from individual studies. The synthesis or simulation method is indicated for each entry. Multiple values from the same reference reflect differing experimental conditions (e.g., annealing temperatures); full details are given in the original works. An extended table including glass phases is provided in the Supporting Information.}
\label{tab:task3_ref}
\end{table}

We benchmark the predicted $\sigma_{298}$ values only by their magnitude, rather than by drawing a direct comparison to specific experimental or computational references, which are compiled in Table \ref{tab:task3_ref}. This is an intentional design choice -- minor variations in experimental synthesis conditions can strongly alter local structural motifs, which can in turn have great influences on the measured ionic conductivity values \cite{kudu_review_2018}. Moreover, such measurements are typically performed on powder samples, where grain boundaries, defects, and amorphous regions play a role -- features that no simulation suitable for large-scale and routine benchmarking purposes can fully capture. A detailed discussion of the limitations associated with comparing experimental and computational $\sigma_{298}$ values is provided in the Supporting Information. 

Performing a complementary AIMD study at comparable simulation size and timescale would also have been highly expensive, and we refrained from doing so. Instead, Task~3 is intended to evaluate whether an MLIP yields conductivity values within a physically reasonable range. In this way, we consider conductivity values within the range of experiment to AIMD to be acceptable. As points of reference, we highlight in bold in Table \ref{tab:task3_ref} the highest reported experimental conductivity of 17 mS/cm \cite{seino_sulphide_2014}, and a representative AIMD value of 61 mS/cm, computed using the same XC-functional as the DFT labels used in LiPS-25 \cite{chu_insights_2016}.

\subsection*{Task 4: Robustness}
The final task assesses the robustness of each MLIP. While this test is not specific to SSEs, it provides a general evaluation of the stability and reliability of MD simulations driven by the model, which are essential for any downstream application. Here, we deliberately push the models to extreme conditions as a stress test, in order to probe the limits of their stability and predictive capability; we note that these conditions are far outside those relevant for LiPS systems, and the resulting structures may be non-physical. We perform 100~ps NPT simulations on a 1008-atom random-hard-sphere structure generated using \texttt{buildcell} \cite{pickard_ab_2011}, and relaxed in a fixed cell with the corresponding potential. Simulations are carried out across a grid of temperatures (1000–16,000~K) and pressures (10\textsuperscript{6}–10\textsuperscript{12} Pa). Robustness is quantified using two metrics: (i) simulation survival, where a green marker denotes that all three repeats completed 100 ps, pale green indicates partial survival (some, but not all, repeats completed 100~ps), and white denotes complete failure (all three repeats failed); and (ii) the number of close-contact events, defined as the number of frames (sampled at 1 ps intervals) containing interatomic separations $\leq 1$~\AA{}.

\FloatBarrier

\section*{Experiments}

\subsection*{Benchmarking Graph-Based MLIPs}

To demonstrate the utility of the tasks accompanying LiPS-25, we study the role of hyperparameters in MACE \cite{batatia_mace_2023}, one of the current state-of-the-art architectures for MLIPs. We fit and evaluate a series of 29 models with systematically varying hyperparameters, using the first three tasks introduced above. It should be emphasized that this is not a conventional hyperparameter optimization aimed at minimizing the test-set loss alone, but rather a broader investigation into which model settings yield the most robust, physically meaningful performance in the context of SSE modeling.

We conduct a sweep over four key hyperparameters: (i) the radial cutoff; (ii) the number of message-passing layers; (iii) the number of channels -- that is, the multiplicity of node features corresponding to each irreducible representation; (iv) the maximal message equivariance, $L$ -- that is, the highest degree of the \textit{O}(3) irreducible representations included in the hidden node features of the network.
We naively compare hyperparameter values that can be considered to be physically reasonable without requiring detailed domain-knowledge of this system, namely: radial cutoffs between 3--8~\AA{}, 1--5 message-passing layers, 8--256 channels, and equivariance degrees of $L=0$ (``small''), $L=1$ (``medium''), and $L=2$ (``large''). These values correspond to including irreducible representations up to degree $L$, specifically: $L=0$: 0\textit{e}, $L=1$: 0\textit{e} +1\textit{o}, and $L=2$: 0\textit{e} + 1\textit{o} + 2\textit{e}, where the number denotes the degree \textit{l} of the representation, and the letters \textit{e} and \textit{o} indicate even and odd parity under inversion, respectively \cite{geiger_e3nn_2022}. For further description of the MACE architecture, we direct the reader to refs.~\citenum{batatia_mace_2023} and \citenum{batatia_design_2025}. All models were trained using the \texttt{graph-pes} software \cite{gardner_graph-pes_2024}, with all other hyperparameters set to their \texttt{graph-pes} defaults.

\begin{figure*}[t]
    \centering
    \includegraphics[width=\linewidth]{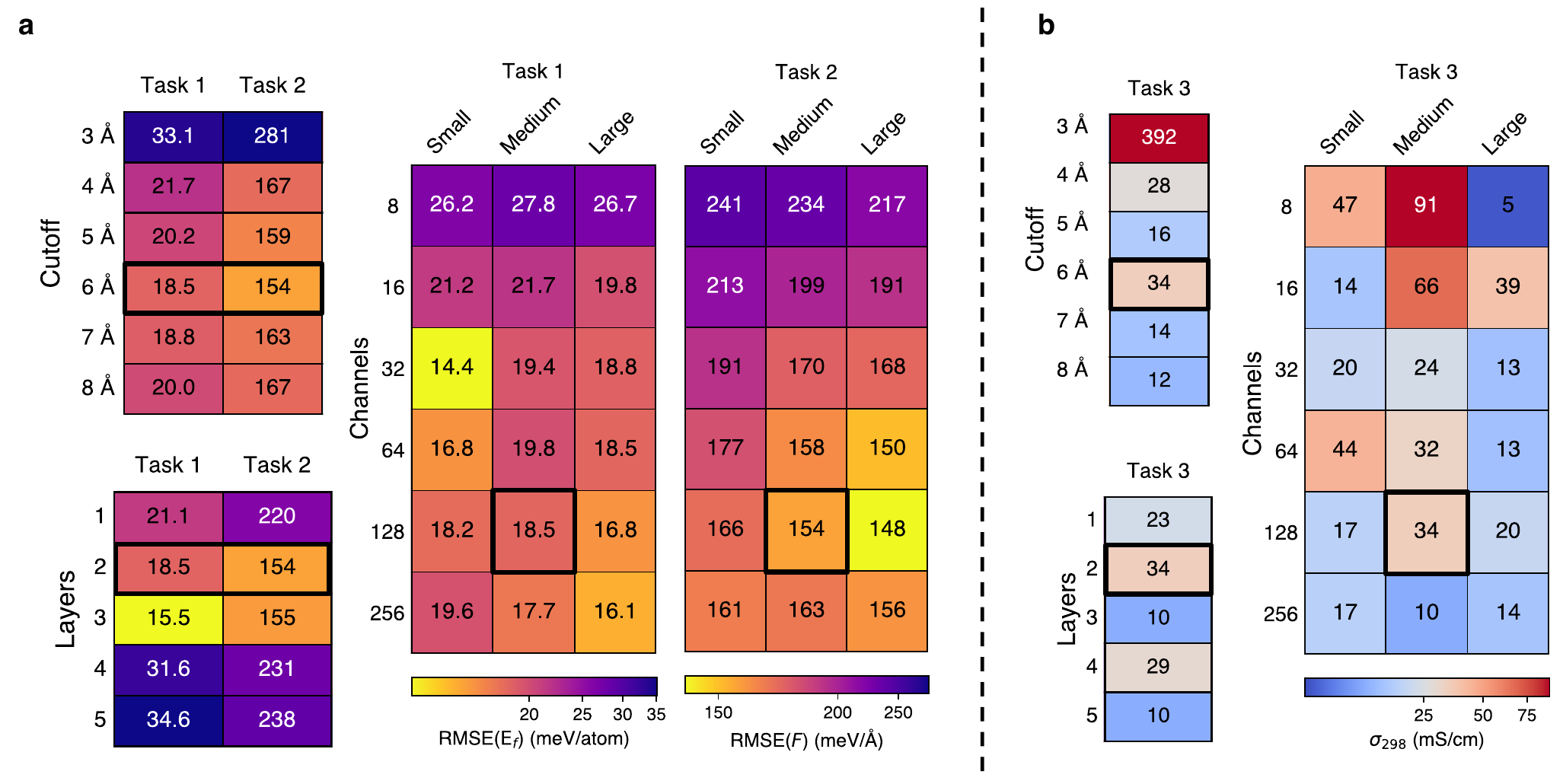}
    \caption{Benchmark task performance for a MACE hyperparameter sweep.
    (a) Performance for formation energies (\textbf{Task 1}) and domain-specific forces (\textbf{Task 2}), reported as the mean of five training repeats. Each value reflects the model's performance under a different hyperparameter configuration, with the metric corresponding to a task-specific prediction error.
    (b) Performance for the predicted magnitude of ionic conductivity, $\sigma_{\text{298}}$, averaged over three repeats (\textbf{Task 3}). As in (a), results reflect variations in model architecture arising from the hyperparameter sweep. 
    In both panels, the boxes outlined in bold indicate the model using a 6~\AA{} cutoff selected from the initial sweep of cutoff radii.
    } 
    \label{fig:figure_3}
\end{figure*}

All trained models were sufficiently stable to perform the MD simulations required for Task 3, enabling the calculation of diffusion coefficients. With this baseline established, we next turn to the influence of individual hyperparameters on predictive accuracy, beginning with the radial cutoff. The performance of MACE models with varying radial cutoffs is characterized in Figure \ref{fig:figure_3}a. Across all tasks, the smallest radial cutoff of 3~\AA{} is strongly penalized -- likely as it fails to capture relevant interactions beyond nearest-neighbor P--S and Li$\cdots$S pairs that are required for accurate energy and force predictions. While the performance across all three tasks stabilizes from a cutoff of 4~\AA{}, and particularly all Task 3 predictions fall within the expected range of conductivity values according to experiment or previous computations (see Table \ref{tab:task3_ref}), minor degradations in Task 1 and 2 errors are seen for cutoff radii larger than 6~\AA{}. Hence, a cutoff of 6~\AA{} (indicated in bold in Fig. \ref{fig:figure_3}) was chosen to be used for subsequent sweeps over layers, channels, and $L$ values. MACE models, like most current MLIPs, are inherently local, and designed to capture short- to medium-range interactions; it is plausible that extending the cutoff to include more distant neighbors introduces noise or redundant information, thereby reducing predictive accuracy.  

Varying the number of message-passing layers has a pronounced impact on performance for Tasks 1 and 2. Single-layer models are too simplistic to capture the LiPS-25 dataset, while deeper architectures (four or five layers) perform worse, possibly due to overfitting or insufficient optimization under the fixed training procedure used here. Two- or three-layer models achieve the best predictive accuracy, suggesting that a moderate depth is expressive enough to capture the system's complexity while remaining transferable within the LiPS-25 domain. In contrast, Task 3 appears to be less sensitive to the number of layers, with all conductivity predictions remaining within the expected range.

In addition to network depth, the choice of model width -- controlled through the number of channels and maximal message equivariance $L$ -- strongly impacts performance. Our results for Tasks 1 and 2 exhibit clear and consistent trends: narrow models (8–16 channels) systematically underperform, while increasing the number of channels leads to reduced errors up to a point, beyond which the improvements plateau. This suggests the existence of an optimal hyperparameter range that captures the relevant underlying physics without introducing unnecessary model complexity. Notably, in Task 1, which relies on both energy and force accuracy in the crystalline domain, invariant (``small'') models perform competitively, and often outperform their equivariant (``medium''/``large'') counterparts. In contrast, our tests for Task 2, which probes force accuracy across diverse environments, demonstrate the advantages of equivariant architectures for predicting vector quantities such as atomic forces. These models benefit from directly learning force vectors via rotation-aware message passing, whereas invariant models must approximate forces through the gradient of a scalar energy field -- an approach that becomes increasingly inaccurate in steep-gradient regimes, such as those encountered in these melt--quench trajectories. The trends for Task 3 are less conclusive: while smaller models produce conductivity estimates that approach the limits of physical plausibility, all models with at least 32 channels yield values within the expected range. However, performance beyond this threshold varies without a consistent trend, underscoring the sensitivity of conductivity predictions to architectural choices. This variability highlights the broader challenge of obtaining reliable and transferable conductivity estimates from simulation alone.

\begin{figure*}[htbp]
    \centering
    \includegraphics[width=\linewidth]{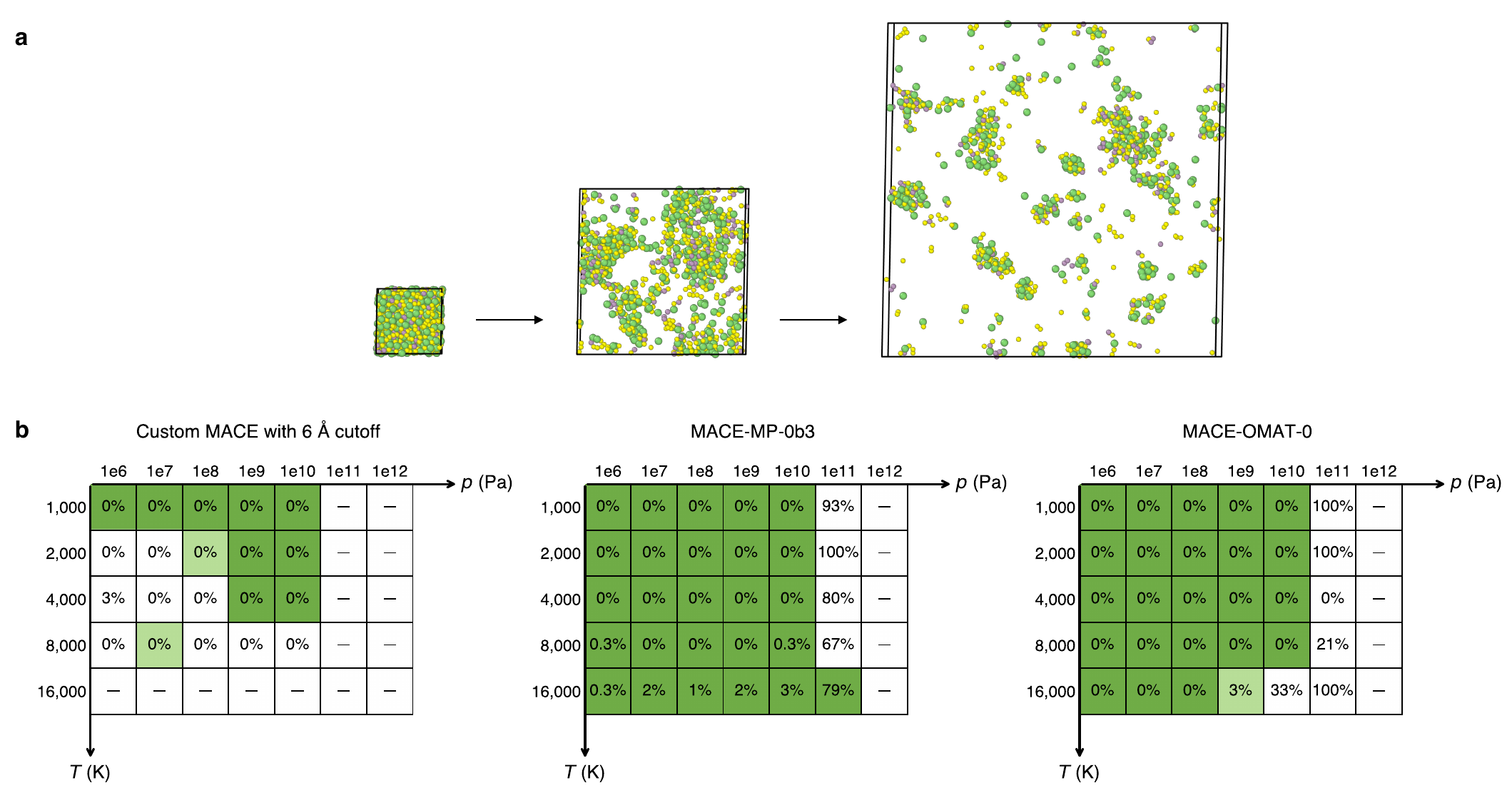}
    \caption{Robustness evaluation of representative MLIP models (\textbf{Task 4}). 
    (a) Example trajectory of an NPT annealing simulation, showing the initial relaxed random-hard-sphere structure and two subsequent configurations illustrating expansion and vaporization under high-temperature / -pressure conditions (atomic color coding: Li, green; P, purple; S, yellow). All structures are drawn to scale and were visualized with OVITO \cite{stukowski_visualization_2009}.
    (b) Grid-search results for three exemplar models: a from-scratch MACE model trained with a 6~\AA{} radial cutoff (from the hyperparameter sweep in Fig.~\ref{fig:figure_3}), and the foundation models MACE-MP-0b3 and MACE-OMAT-0. For each model, 100 ps NPT anneals were performed across a grid of temperatures (1000–16,000~K) and pressures ($10^{6}$–$10^{12}$~Pa) with three repeats. Boxes are shaded green if all repeats reached 100 ps, pale green if one or two repeats reached 100 ps, and white if all repeats failed. The percentages inside the boxes denote the fraction of frames (sampled every 1 ps) with interatomic separations of $\leq 1$~\AA{}. A dash (``--'') indicates that all three repeats failed before 1 ps, i.e., no frames were available for evaluation.
    } 
    \label{fig:figure_4}
\end{figure*}

As a final and complementary benchmarking study, we investigate the robustness of one MACE model trained from scratch -- specifically the 6~\AA{} cutoff variant highlighted in bold in Figure \ref{fig:figure_3} -- and two foundation models, MACE-MP-0b3 and MACE-OMAT-0, using Task 4. The results are shown in Figure \ref{fig:figure_4}. Robustness is quantified using two metrics: (i) the survival of the simulation to 100 ps, and (ii) the number of close-contact events, defined as interatomic separations $\leq$ 1~\AA{}. 

Both foundation models clearly outperform the from-scratch MACE model across these criteria. This is expected, since LiPS-25 is a comparatively narrow training domain, comprising only $\sim$13k structures sampled up to 2000~K and at ambient pressure (1 atm) during melt--quench iterations. In this sense, the from-scratch model is in fact demonstrating robustness beyond its training domain, with successful simulations observed at temperatures up to 4000~K and pressures up to $10^{10}$~Pa. In contrast, the foundation models are trained on vastly larger datasets (1.6M structures for MACE-MP-0b3 and 101M structures for MACE-OMAT-0), which encompass a broader range of chemical and structural environments.

Interestingly, MACE-MP-0b3 appears robust over a slightly wider range of $(T,p)$ combinations than MACE-OMAT-0, despite being trained on a dataset that is two orders of magnitude smaller. While the OMAT dataset \cite{barroso-luque_open_2024} was specifically designed to extend the Alexandria dataset \cite{schmidt_improving_2024} with additional off-equilibrium configurations, this larger and more diverse dataset does not appear to provide significant stability benefits in extreme temperature--pressure conditions within this specific task.

The close-contact metric proves to be well correlated with simulation survival. For both the from-scratch MACE model and MACE-OMAT-0, simulations that fully survive (marked in green, indicating all three repeats reached 100 ps) invariably contain 0\% close-contact frames, whereas partial-survival (in pale green) and failure (in white) regions show progressively higher fractions of such frames. MACE-MP-0b3, however, deviates from this trend: some fully surviving trajectories, such as that at 16,000~K and $10^{11}$~Pa, contain extremely high fractions of ``bad'' frames (up to 79\%), yet do not fail catastrophically. Taken together, these results demonstrate that while pre-training on large datasets can enhance the stability of MLIPs under extreme conditions, the reliability of the resulting trajectories is not guaranteed and depends on the specific model and simulation regime.

\subsection*{Fine-Tuning Foundation Models}

We further demonstrate the utility of the LiPS-25 dataset by using it to fine-tune atomistic foundation models (FMs). To focus this study, we assess the performance of models fine-tuned specifically for the \ce{Li7P3S11} composition. A subset of approximately 400 \ce{Li7P3S11} structures was extracted from the full LiPS-25 dataset to serve as a fine-tuning dataset. Several current leading FMs, selected based on their performance on benchmarks such as the Matbench leaderboard at the time of experiment design \cite{riebesell_framework_2025}, were fine-tuned using \texttt{graph-pes}. We aim to better understand the effects of both the pretraining dataset and the model architecture on fine-tuning procedures. We first compare versions of MACE FMs \cite{batatia_foundation_2025} (namely, MACE-MP-0b3, MACE-MPA-0, MACE-OMAT-0, and MACE-MATPES-PBE-0), which share largely the same architectures and therefore primarily reflect differences in the pretraining dataset. We then extend this comparison to the MatterSim \cite{yang_mattersim_2024} (both MatterSim-1m and MatterSim-5m) and Orb \cite{neumann_orb_2024, rhodes_orb-v3_2025} (namely, Orb-v2, Orb-v3-direct-inf-mpa, Orb-v3-direct-inf-omat) families, where architectural differences also play a significant role. Their performance is evaluated in an extended version of Task 2, incorporating both force and energy accuracy tests, as well as on Task 3. To account for differences in reference atomic energies between the pretraining datasets and LiPS-25 arising from variations in exchange--correlation functionals and pseudopotentials, the \texttt{add\_auto\_offset} feature of \texttt{graph-pes} was applied to correct the zero-shot energy predictions with an offset, thereby aligning them with the energy scale of LiPS-25.

\begin{figure*}[htbp]
    \centering
    \includegraphics[width=0.9\linewidth]{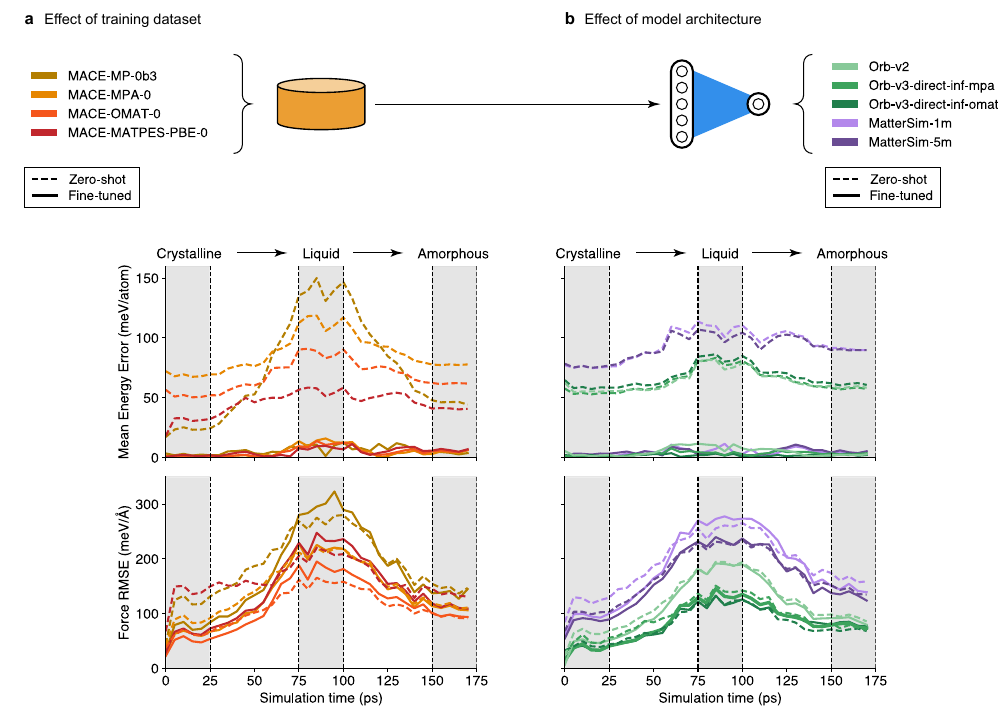}
    \caption{Accuracy of foundation models fine-tuned on 25 \ce{Li7P3S11} structures from LiPS-25. 
    We assess these models on a \ce{Li7P3S11} melt--quench trajectory (Task 2), showing energy ({\em top}) and force ({\em bottom}) errors against DFT-labeled snapshots. Errors for fine-tuned models (solid lines) are averaged over 5 models fine-tuned with different seeds. Zero-shot errors (dashed lines), corresponding to models evaluated without fine-tuning, are shown for comparison. For these zero-shot models, energy predictions were corrected using the \texttt{add\_auto\_offset} feature of \texttt{graph-pes} to account for differences in reference atomic energies between the pretraining datasets and LiPS-25, arising from the use of different exchange--correlation functionals and pseudopotentials. 
    (a) Schematic for an atomistic ML model, showing the mapping between the dataset and the model architecture. (b) Performance of MACE foundation models: assessing the effect of differences in training dataset for similar architecture. (c) Performance of other foundation model families, viz.\ MatterSim and Orb: assessing differences in model architecture.}
    \label{fig:figure_5}
\end{figure*}

Figure \ref{fig:figure_5} compares several fine-tuned FMs from the MACE, MatterSim, and Orb families. All models shown have been fine-tuned on the same 25 structures, randomly selected from the filtered \ce{Li7P3S11} dataset; preliminary investigations demonstrated that performance saturates beyond 25 fine-tuning structures (Figure S3a). Across all FMs, it is evident that fine-tuning has a stronger effect on energy errors than force errors (despite having pre-corrected for effects of different DFT functionals). A possible explanation could be that whilst fine-tuning shifts the relative positions of minima on the potential-energy surface, it maintains relatively similar gradients between them. 

Clear trends emerge in model performance with respect to the choice of pretraining dataset. Both the MACE and Orb families exhibit consistent improvements in energy and force errors as the pretraining dataset progresses from MPTrj\cite{deng_chgnet_2023} to MPA \cite{schmidt_improving_2024} to OMat24 \cite{barroso-luque_open_2024} -- although this effect is notably more pronounced for MACE models. The MPA dataset introduces additional structural diversity through the inclusion of the sAlex dataset, complementing the DFT-relaxed frames of the MPTrj baseline. This enables models such as MACE-MPA-0 to improve upon their MPTrj-pretrained counterparts, like MACE-MP-0b3. Models pre-trained on OMat24 consistently perform best across both the MACE and Orb families. We attribute this advantage to OMat24's emphasis on non-equilibrium structures, generated by applying perturbations, such as rattling and AIMD, to configurations from the Alexandria dataset. We think that this proves especially beneficial in out-of-equilibrium regimes, such as the liquid state explored here, where the difference between OMat24 and other datasets becomes more pronounced. Notably, the MATPES dataset \cite{kaplan_foundational_2025}, comprising 400k frames from MD trajectories of MP structures, yields a MACE model with performance comparable to models trained on much larger datasets (1.6M MPTrj, 12M MPA, 101M OMAT24). This underscores the value of judiciously sampled data, and suggests that impactful model development remains feasible even with more modest computational resources.

The MatterSim models were pre-trained on a distinct dataset comprising structures from the MP, Alexandria, the ICSD, and internally generated configurations \cite{yang_mattersim_2024}. Whilst increased architectural complexity (from 1m to the 5m version) improves their performance, these models still exhibit higher errors in this task relative to most MACE and all Orb FMs assessed here. This may reflect their comparatively smaller size -- even the largest 5M-parameter MatterSim model is notably smaller than other pre-trained models such as MACE-MPA-0 (9M parameters) and Orb-v3 (25M parameters) -- or other architectural differences between frameworks. Further analysis would be required to clarify the respective roles of model capacity and architecture. 

The most accurate model overall in this benchmark is Orb-v3-direct-inf-omat, which benefits both from pre-training on the high-quality OMat24 dataset and from architectural advantages shared across the Orb family. In particular, its use of direct force prediction, rather than inferring forces from energy gradients, likely contributes to its higher accuracy. However, the numerical gains and efficiency boost of such non-conservative models must be weighed against drawbacks such as poorly converged optimizations and inaccuracies in MD, particularly for collective processes involving long-range correlations, as recently discussed by Bigi et al. (ref. \citenum{bigi_dark_2025}). These limitations may particularly influence $\sigma_{298}$ predictions in the present study.

Perhaps most striking, rather than individual model performance, is the collective domain-specific performance of fine-tuned FMs. Within the crystalline regime, fine-tuned FMs can improve force errors upon their zero-shot counterparts by up to a factor of two. For amorphous structures, the benefit is smaller but often still observable. However, in the liquid state, the fine-tuned models are consistently outperformed by their zero-shot analogues, despite the fact that the fine-tuning dataset includes liquid \ce{Li7P3S11} structures. This suggests that fine-tuning on a mixed-domain dataset can result in loss of knowledge in cases where structural disorder is pronounced. Notably, MACE-MP-0b3 exhibits catastrophic forgetting in the liquid regime, meaning that fine-tuning causes the model to lose previously learned behavior, as evidenced by a marked change in its error profile compared with the zero-shot model. In contrast, the other fine-tuned models retain the same error profile shape as their zero-shot counterparts, albeit with systematically higher errors. Orb models are an exception here -- models either reproduce zero-shot behavior or offer small improvements in the liquid regime. These findings highlight a central challenge: fine-tuning can have markedly different effects on model performance in regimes of different degrees of structural disorder, with improvements in one regime sometimes coming at the expense of another. Future work should therefore focus on designing fine-tuning procedures that preserve foundation models’ generalizability and cross-domain robustness, which is essential for realizing their advantages over conventional task-specific models.

\subsection*{Ionic Conductivities from Fine-Tuned Models}

\begin{figure*}[ht]
    \centering
    \includegraphics[width=\linewidth]{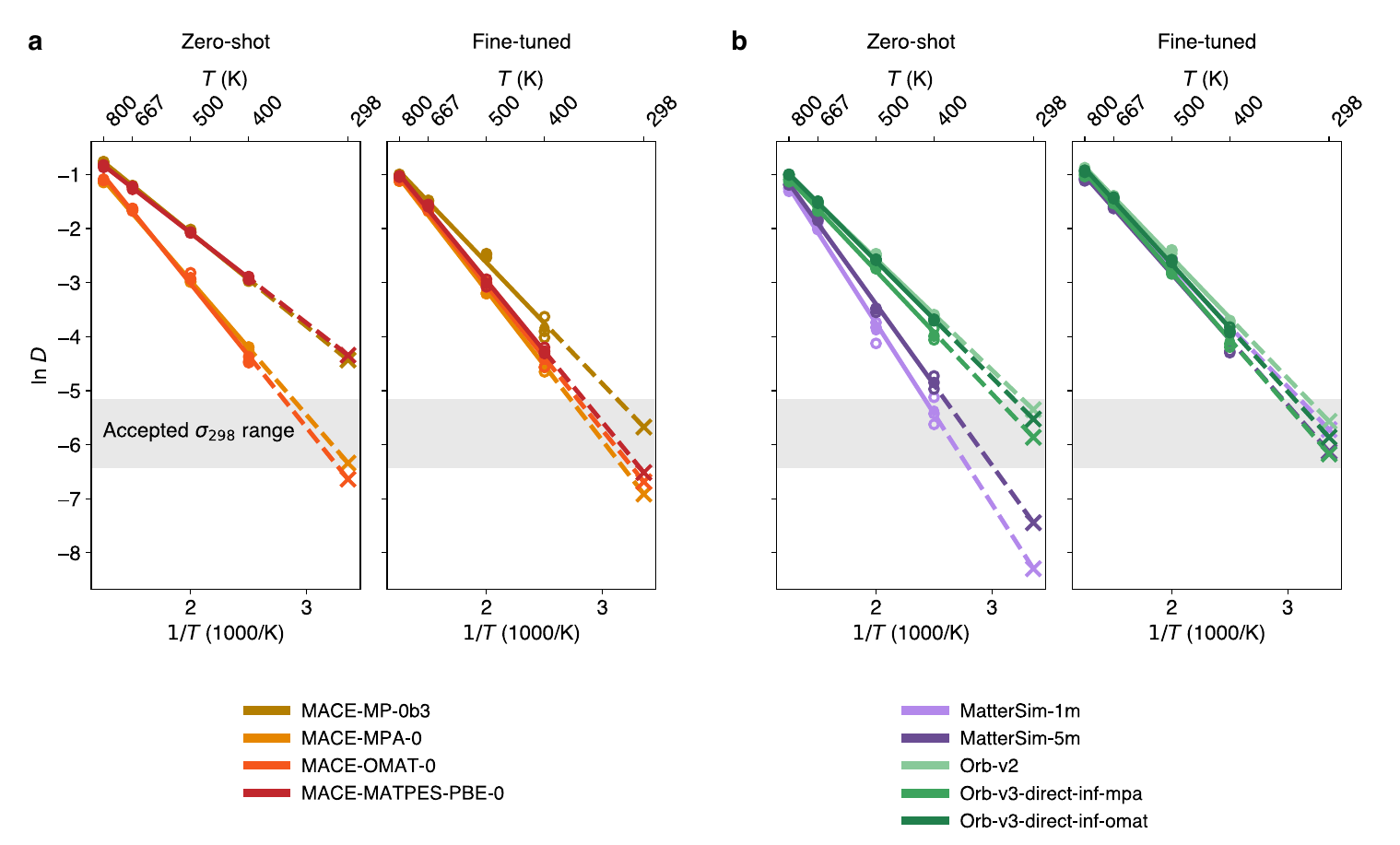}
    \caption{Effect of fine-tuning on room-temperature ionic conductivity predictions for crystalline \ce{Li7P3S11}. 
    (a) Arrhenius plots for MACE foundation models. 
    (b) Same for for MatterSim and Orb models. 
    In both cases, zero-shot predictions ({\em left}) are compared to predictions by models that have been fine-tuned on 25 randomly selected \ce{Li7P3S11} structures from the LiPS-25 dataset ({\em right}). 
    Hollow circles represent \(D(T)\) values from individual trajectories; filled circles indicate mean \(D(T)\) across three repeats. Solid lines show the best fit to the mean values; dotted lines represent extrapolation to \(D(298\,\mathrm{K})\).
    The grey shaded region indicates the region of accepted $\sigma_{\text{298}}$ values according to the literature, as described in the main text.}
    \label{fig:figure_6}
\end{figure*}

As a final assessment of the fine-tuned FMs, we now proceed to Task 3 to evaluate their performance in predicting the room-temperature ionic conductivity of \ce{Li7P3S11}. While all zero-shot models exhibit qualitatively correct linear Arrhenius behavior (left-hand-side panels of Figures \ref{fig:figure_6}a and \ref{fig:figure_6}b), the corresponding extrapolated ionic conductivities at 298~K vary significantly between models, even by more than an order of magnitude within the same model family (see MACE models in Table \ref{tab:figure_6_table}). Notably, models such as MACE-MP-0 and MACE-MATPES-PBE-0 overpredict ionic conductivity relative to the expected range (see Table \ref{tab:task3_ref}), consistent with a systematic softening of the underlying PES. This behavior has previously been attributed to pretraining datasets biased towards near-equilibrium configurations, typically derived from DFT relaxation trajectories \cite{deng_systematic_2025}. In contrast, models pre-trained on more structurally diverse datasets, such as MPA or OMat24, consistently produce zero-shot conductivity predictions within a physically reasonable range, regardless of architecture. This mirrors the trends observed in the domain-specific errors of Task 2 (Figure \ref{fig:figure_5}), where the same models exhibited lower errors in high-temperature configurations along the melt--quench trajectory. Such results suggest that strong performance in domain-specific error benchmarks, as in Task 2, can be a useful indicator of reasonable dynamic performance. MatterSim models, on the other hand, tend to underpredict conductivity, which could indicate an overly rigid PES that suppresses ion mobility; here, both pretraining coverage and architectural differences likely contribute to the observed trends.

\begin{table}[t]
    \caption{Predicted ionic conductivities of \ce{Li7P3S11} at 298 K before and after fine-tuning. The reported $\sigma_{\text{298}}$ values are calculated from the average diffusion coefficients, \(D(T)\), obtained across three independent repeats (corresponding to the filled-circle data and fitted lines in Figure \ref{fig:figure_6}). All values are rounded to the nearest integer.}
    \centering
    \begin{tabular}{lcc}
        \hline
          & \multicolumn{2}{c}{$\sigma_{\text{298}}$ (mS/cm)} \\
        \cline{2-3}
        \textbf{Model} & \textbf{Zero-shot} & \textbf{Fine-tuned}
        \\ \hline
        MACE-MP-0b3   & 125 & 36  \\ 
        MACE-MPA-0   & 18 & 11  \\ 
        MACE-OMAT-0   & 14 & 13 \\ 
        MACE-MATPES-PBE-0   & 137 & 16  \\ 
        \hline
        MatterSim-1m   & 3 & 35 \\ 
        MatterSim-5m   & 6 & 23  \\
        Orb-v2 & 50 & 40  \\
        Orb-v3-inf-direct-mpa & 30 & 22  \\
        Orb-v3-direct-inf-omat & 42 & 30  \\
        \hline
    \end{tabular}
    \label{tab:figure_6_table}
\end{table}

Fine-tuning the models on the same subset of 25 \ce{Li7P3S11} structures as in Fig. \ref{fig:figure_5} brings their predictions into much closer agreement, both across and within model families. This convergence is evident in the right-hand-side panels of \ref{fig:figure_6}a and \ref{fig:figure_6}b, where the Arrhenius lines of fine-tuned models align more closely. Moreover, all fine-tuned models yield conductivity values within the expected range (see Table \ref{tab:task3_ref}). Nonetheless, systematic differences between families persist: for example, fine-tuned MACE models consistently predict lower ionic conductivity values than their MatterSim and Orb counterparts (with the exception of MACE-MP-0b3), despite improved overall agreement. These findings highlight the importance of fine-tuning in correcting systematic biases inherited from pre-training, while also suggesting that residual architectural or training differences between model families continue to influence dynamic properties such as ionic conductivity.

Looking ahead, such validation of MLIPs designed for complex functional materials on the physical properties they aim to reproduce should become standard practice. For ionic conductors, this involves assessing transport behavior like ionic conductivity, despite the associated conceptual and practical challenges. Since conductivity predictions are computationally demanding, it is useful to first apply more affordable, targeted tests, such as the domain-specific error analysis as in Task 2, which can serve as strong indicators of dynamic performance. The lack of a clear ground truth complicates validation further: experimental values are not directly comparable to simulations, and generating fully converged AIMD references for each target system would be impractical. As such, future benchmarking efforts should integrate property-level evaluations with complementary analyses, such as inspecting diffusion mechanisms or jump statistics, to ensure that predicted transport arises from physically reasonable processes, even in the absence of an exact conductivity reference.

\clearpage

\section*{Conclusions and Outlook}

As machine-learning acceleration becomes the norm in computational materials chemistry, the careful and systematic evaluation of MLIP models is ever more important. The Li--P--S system is well-suited for this purpose: both because of the inherent interest in the materials themselves, and because it represents a broader class of complex chemistries and dynamic phenomena with relevance to battery research.
Our LiPS-25 dataset supports the fitting of MLIPs for Li--P--S materials and, perhaps even more importantly, the benchmarking of existing and new models. We have shown examples of how LiPS-25 can enable insights into the nature and applicability of graph-based foundation MLIPs, as well as into fine-tuning strategies.

Looking forward, physically grounded benchmarks like those presented here can serve as a general template for validating MLIPs. We have outlined protocols that span four levels of evaluation: starting with basic energetic validation (Task 1) and domain-specific force accuracy tests along relevant MD trajectories (Task 2) through to full dynamic property benchmarks, here, the ionic conductivity (Task 3), and finally to an assessment of robustness under a wide range of conditions, including very high temperatures and pressures (Task 4). Together, these tasks provide a structured framework for assessing MLIP quality that can guide model developers and users. While we have focused on the Li--P--S system in the present study, we expect that the framework (and associated code) can be readily adapted to other material systems of interest.

In the age of atomistic foundation models, systematic tests as outlined in this work could be incorporated into validation pipelines \cite{zills_mlipx_2025}, community benchmarks \cite{dunn_benchmarking_2020, riebesell_framework_2025}, and automated MLIP development workflows \cite{janssen_pyiron_2019, liu_automated_2025}. Embedding LiPS-25 and related benchmarks in this way would not only clarify how models behave across different regimes, but also guide their most effective use in downstream applications -- ultimately supporting more reliable, transparent, and efficient use of MLIPs in computational materials chemistry.

\clearpage

\section*{Data availability}

Data supporting this work are available at \url{https://github.com/nfragapane/lips-25}.

\begin{acknowledgement}

We thank Chiheb Ben Mahmoud for helpful discussions and Han Yang for helpful comments on the manuscript. This work was supported by UK Research and Innovation [grant number EP/X016188/1]. We are grateful for computational support from the UK national high performance computing service, ARCHER2, for which access was obtained via the UKCP consortium and funded by EPSRC grant ref EP/X035891/1 (see also ref.~\citenum{beckett_archer2_2024}). We are grateful to the UK Materials and Molecular Modelling Hub for computational resources, which is partially funded by EPSRC (EP/T022213/1, EP/W032260/1 and EP/P020194/1).

\end{acknowledgement}

\setstretch{1.0}

\providecommand{\latin}[1]{#1}
\makeatletter
\providecommand{\doi}
  {\begingroup\let\do\@makeother\dospecials
  \catcode`\{=1 \catcode`\}=2 \doi@aux}
\providecommand{\doi@aux}[1]{\endgroup\texttt{#1}}
\makeatother
\providecommand*\mcitethebibliography{\thebibliography}
\csname @ifundefined\endcsname{endmcitethebibliography}  {\let\endmcitethebibliography\endthebibliography}{}

\end{document}

\maketitle

\clearpage

\setstretch{1.1}

\section{Dataset Construction}
\subsection{Initial Dataset (``Iter0'')}
The LiPS-25 dataset is constructed to focus on the pseudo-binary \ce{Li2S}–\ce{P2S5} tie-line while maintaining broad coverage of the Li--P--S configurational space. The initial dataset (Iter0) and subsequent iterative melt–quench augmentations (Iter1-\textit{x}, Iter2-\textit{x}) were built around seven key compositions (see Figure 1a of the main text). The specific crystal structures used as starting points for AIMD annealing or melt–quench simulations are as follows: the constituent binary phases, \ce{Li2S} (anti-fluorite, $\mathit{Fm}\overline{3}\mathit{m}$; ICSD 60432), and \ce{P2S5} (a regular arrangement of \ce{P4S10} molecules, $\mathit{P}\overline{1}$; ICSD 409061), as well as relevant ternary compounds, viz.\ \ce{Li2P2S6} ($\mathit{C}2/\mathit{m}$; ICSD 253894), \ce{Li4P2S7} ($\mathit{P}\overline{1}$; ref.~\citenum{holzwarth_computer_2011}), \ce{Li7P3S11} ($\mathit{P}\overline{1}$; ICSD 157654), \ce{Li3PS4} ($\mathit{Pmn}2_1$; ICSD 180318), and \ce{Li7PS6} ($\mathit{Pna}2_1$; mp-1211324).

The Iter0 dataset comprises the following components, which are detailed below beyond the description given in the main text:
\begin{itemize}
\item \textit{Crystalline}: All elemental (Li, P, S), binary, and ternary crystalline structures listed in the ICSD \cite{zagorac_recent_2019} and Materials Project \cite{jain_commentary_2013} were included, with duplicates between the two databases removed and entries without full site occupancy excluded. For ternary crystals, both the primitive unit cells and $2 \times 2 \times 2$ supercells were considered. These structures underwent an initial DFT relaxation, followed by either a volume ($\pm 10\%$ around the relaxed volume) or angle (random angles within a range of 20\% of relaxed cell angles) distortion, and atomic position ``rattling'' (with a standard deviation of 0.01\textup{~\AA}). These distortions aim to provide sampling around local minima of the potential energy surface. 

\item \textit{AIMD snapshots}: For each of the seven key crystal structures, three separate 20~ps NVT AIMD runs were performed at 250, 500, and 1000~K, each at four densities scaled between the relaxed density and $2 \times$ relaxed density. Every 1000-th frame was extracted and labeled.

\item \textit{Random Hard Sphere (RHS) Models}: Structures were generated using the \texttt{buildcell} code of \textit{ab initio} random structure searching (AIRSS) \cite{pickard_ab_2011}, with the latter code accessed using the Autoplex \cite{liu_automated_2025} package. A minimum separation between atom pairs was enforced, defined as the average experimental crystalline values minus 0.5 \textup{\AA{}}. 

\item \textit{Dimers}: Dimer configurations of all Li, P, and S pairs (viz.\ Li--Li, P--P, S--S, Li--S, Li--P, P--S) were sampled with interatomic separations of 1.0--2.0 \textup{\AA{}} in 0.1 \textup{\AA{}} intervals, and 2.0--7.0 \textup{~\AA} with 0.2 \textup{\AA{}} intervals, within 20 $\times$ 20 $\times$ 20~\AA\ boxes. These configurations provide reference data for isolated pair interactions, including short-range repulsions and the onset of longer-range attractions.

\end{itemize}

\subsection{Iterative Training (``Iter1'' and ``Iter2'')}
NequIP-driven \cite{batzner_e3-equivariant_2022} melt–quench (MQ) simulations were employed to iteratively expand the Iter0 dataset, starting from the seven key compositions described above. At each iteration, the most uncertain structures were identified using a query-by-committee procedure: five subsampled models, each trained on a random 50\% of the current dataset, were used to make predictions on a pool of evenly spaced snapshots collected from the MQ trajectories of all seven compositions. The 250 structures with the largest standard deviation in force predictions were then labeled and added to the dataset. 

A timestep of 1~fs was used for all simulations. NVT runs employed a thermostat damping constant \(t_{\text{damp}}^{(T)} = 100\)~fs, while NPT runs used \(t_{\text{damp}}^{(T)} = 10\)~fs and a barostat damping constant \(t_{\text{damp}}^{(p)} = 100\)~fs. Complete MQ simulation protocols for each iteration are provided in Table~\ref{tab:mq_protocols}.

\begin{table}[t]
    \centering
    \begin{tabular}{ccccc}
        \hline
        & & \textbf{MQ Temperatures} & \textbf{Quench Rate} & \\
        \textbf{Iteration} & \textbf{Ensemble} & \textbf{(K)} & \textbf{(K/ps)} & \textbf{Structure type}\\ \hline
        Iter1-1   & NVT   & 300-1000-300  & 50 & All\\ \hline
        Iter1-2   & NVT   & 300-1000-300  & 50 & All\\ \hline
        Iter1-3   & NVT   & 300-1500-300  & 50 & All\\ \hline
        Iter1-4   & NVT   & 300-1500-300  & 100 & All\\ \hline
        Iter2-1   & NPT   & 300-1500-300  & 50 & Glasses only\\ \hline
        Iter2-2   & NPT   & 300-1500-400  & 50 & Glasses only\\ \hline
        Iter2-3   & NPT   & 300-1500-500  & 50 & Glasses only\\ \hline
    \end{tabular}
    \caption{MQ protocols used in the iterative training procedure. The table reports the ensemble, temperature range for the melt-quench in the format ``$T_\text{start}$-$T_\text{melt}$-$T_\text{quench}$'', quench rate, and the type of structure extracted by the query-by-committee (QbC) procedure. The label ``All'' indicates that QbC had unrestricted selection of the most uncertain structures across compositions and trajectory points, including disordered crystalline, liquid, and glassy states. ``Glasses only'' indicates that QbC was restricted to the annealing stage following MQ, such that only amorphous structures were included.}
    \label{tab:mq_protocols}
\end{table}

\clearpage

\section{Benchmark Tasks}
\subsection{Task 1: Energetic accuracy}
\subsubsection{Starting structures}
For this task, the formation energies of eight relevant structures along the tie-line were calculated: \ce{Li2P2S6} (ICSD 253894), \ce{Li4P2S7} (ref.~\citenum{holzwarth_computer_2011}), \ce{Li7P3S11} (ICSD 157654), $\alpha$-\ce{Li3PS4} (ref.~\citenum{homma_crystal_2011}), $\beta$-\ce{Li3PS4} (mp-985583), $\gamma$-\ce{Li3PS4} (ICSD 180318), low-temperature \ce{Li7PS6} ($\mathit{Pna}2_1$, mp-1211324), and high-temperature \ce{Li7PS6} ($\mathit{F}\overline{4}3\mathit{m}$, ICSD 421130; partial occupancies were resolved using the \texttt{supercell} program \cite{okhotnikov_supercell_2016}).

\subsubsection{Alternative calculation details and results}
\begin{figure*}[h]
    \centering
    \includegraphics[width=0.5\linewidth]{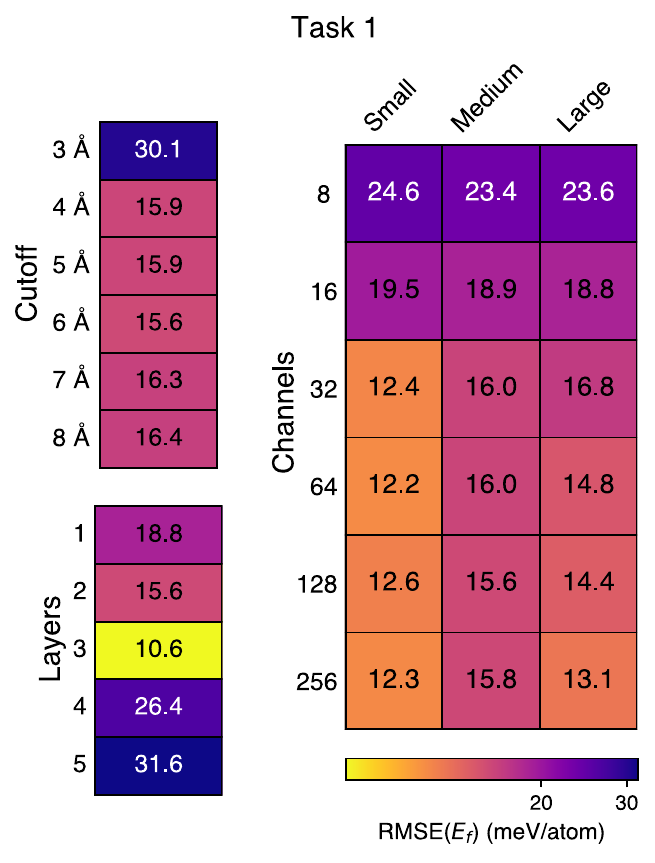}
    \caption{Task-1 performance for a MACE hyperparameter sweep. Errors are computed using the alternative protocol to the main text, i.e., from a fixed set of DFT-relaxed structures, and reported as the mean over five training repeats.
    } 
    \label{fig:figure_1}
\end{figure*}

In the task as described in the main text, RMSE($E_{\text{f}}$) is evaluated by relaxing each structure with either DFT or the MLIP model and then labeling with the same method -- in line with the formal definition of $E_{\text{f}}$, but thereby conflating energetic and force accuracies. Here, we also provide the RMSE($E_{\text{f}}$) computed from a fixed set of DFT-relaxed structures to isolate single-point energetic accuracy. These results are shown in Figure \ref{fig:figure_1}, and both protocols are implemented in the Jupyter notebook accompanying the present work.

\subsection{Task 2: Domain-specific force accuracy}
\subsubsection{Simulation details}
A melt–quench simulation of a 672-atom \ce{Li7P3S11} supercell was performed in the NPT ensemble between 300 and 1500~K using a 1~fs timestep. The protocol consisted of a 25~ps annealing run at 300~K, a 50~ps melt ramp to 1500~K, a 25~ps anneal at 1500~K, and a 25~ps quench back to 300~K (corresponding to melt and quench rates of 24~K/ps),with damping parameters \(t_{\text{damp}}^{(T)} = 10\)~fs and \(t_{\text{damp}}^{(p)} = 100\)~fs. The simulation was driven by an interim NequIP potential from Iter2-2. DFT snapshots were extracted every 5~ps along the trajectory, and MLIP force errors were evaluated against these labels. The corresponding LAMMPS\cite{thompson_lammps_2022} scripts  and Jupyter Notebook for analysis are provided.

\subsection{Task 3: Property accuracy}
\subsubsection{Simulation details}
500~ps NVT anneals were performed on 672-atom \ce{Li7P3S11} supercells at 400, 500, 667, and 800~K, with a timestep of 1~fs and \(t_{\text{damp}}^{(T)} = 100\)~fs. Example LAMMPS \cite{thompson_lammps_2022} scripts are provided.

\subsubsection{Ionic conductivity calculation details}
The Li-ion mean-square-displacement (MSD) is extracted from MD simulations using the MDAnalysis package \cite{michaud-agrawal_mdanalysis_2011, gowers_mdanalysis_2016}, according to:
\begin{equation}
    \text{MSD}(t) = \frac{1}{N_{\text{Li}}} \sum_{i=1}^{N_{\text{Li}}} \left[ \mathbf{R}_i(t) - \mathbf{R}_i(t=0) \right]^2
    \label{eq:msd}
\end{equation}
where \textit{N}$_{\text{Li}}$ is the total number of Li ions, and \textbf{R}$_{i}$ is the position of the \textit{i}-th Li ion. 

The diffusion coefficient ($D_{\text{Li}}$) at each temperature is extracted from the slope of a linear fit to MSD vs \textit{t} per block:
\begin{equation}
    D_{\text{Li}} = \frac{\text{MSD}(t)}{2 d t}
    \label{eq:diffusion_coeff}
\end{equation}
where $d$ is the diffusion dimension ($d=3$ here). Analysis of Li-ion motion is restricted to only the linear regime of MSD vs \textit{t}, excluding the initial ballistic regime (the first 10~ps of the trajectory), and the block-averaging method (averaging $D_{\text{Li}}$ over blocks of 20~ps) is used to extract a mean $D_{\text{Li}}$ value.  

The Arrhenius relation can be fitted to the temperature-dependent $D$ values:
\begin{equation}
    D_{\text{Li}}(T) = D_0 \exp\left( \frac{-E_{\text{a}}}{k_{\text{B}} T} \right)
    \label{eq:arrhenius}
\end{equation}
where $D_{0}$ and E$_{\text{a}}$ refer to the pre-exponential factor and the activation energy, respectively, and the Boltzmann factor is given by $k_{\text{B}}$. These same values can be used to extrapolate to $D$ at 298 K. 
To estimate the ionic conductivity at \textit{T}=298 K, $\sigma_{\text{298K}}$, the Nernst--Einstein relation is then used:
\begin{equation}
    \sigma = \frac{N_{\text{Li}} q^2 D_{\text{Li}}(T)}{V k_{\text{B}} T}
    \label{eq:nernst_einstein}
\end{equation}
where $V$ is the total volume of the simulated system, and \textit{q} is the ionic charge of the Li$^{+}$ charge carriers (i.e., $q=e$). The accompanying Jupyter notebook for such trajectory analysis is provided.

It is noted that there are potential limitations associated with using the Nernst--Einstein relation in SSEs, particularly when there exist dynamical correlations between the Li ion charge carriers, and between these charge carriers and the thiophosphate backbone. Correlations between carriers allow for a cooperative motion that can artificially increase the ionic conductivity prediction.  A more accurate alternative would consider center-of-mass motion to calculate the charge diffusion coefficient \cite{marcolongo_ionic_2017}; however, to achieve the statistical accuracy required for this approach, much longer simulations would need to be run that are not feasible for the purpose of benchmarking.

\begin{figure*}[h]
    \centering
    \includegraphics[width=\linewidth]{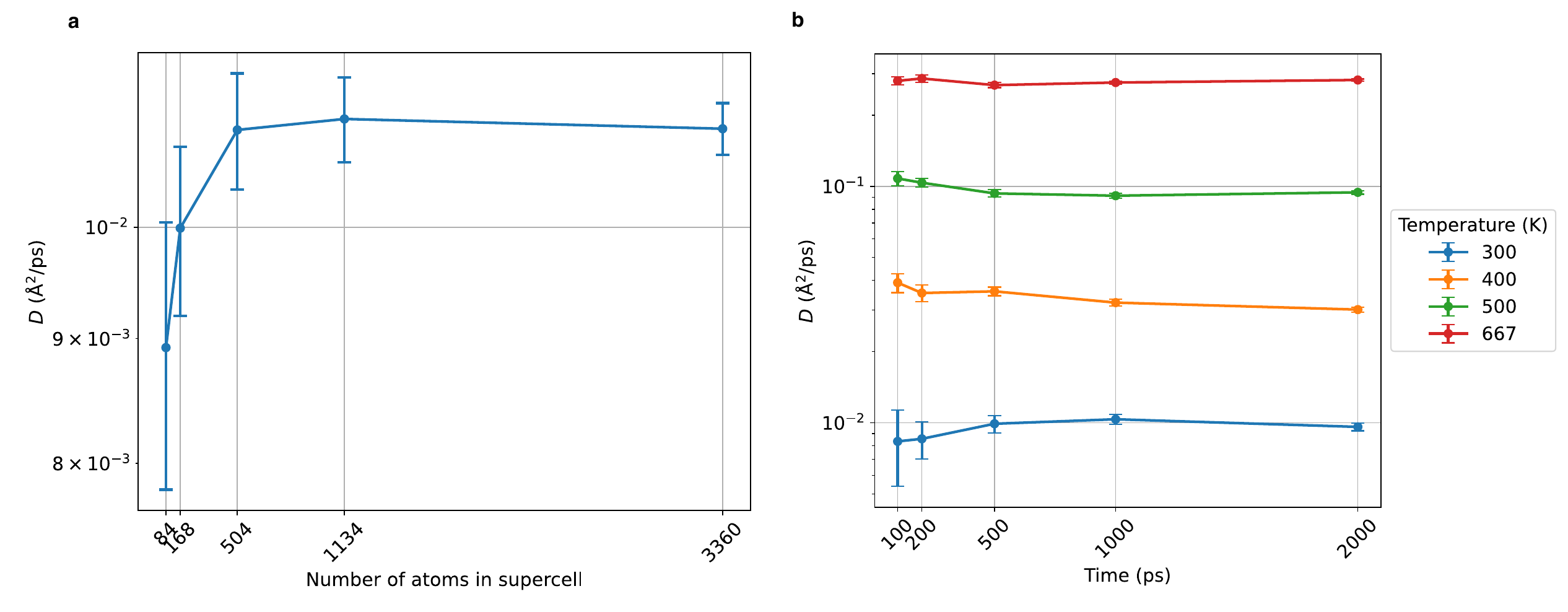}
    \caption{Convergence of predicted diffusion coefficients, $D$, with respect to (a) supercell size and (b) simulation length. In (a), values are obtained from 1~ns trajectories at 300~K for \ce{Li7P3S11} supercells of increasing size. In (b), diffusion coefficients are evaluated for simulation lengths up to 1~ns at temperatures between 300 and 667~K using a 672-atom \ce{Li7P3S11} supercell. Error bars represent the block-averaged standard error, i.e., the standard deviation of block diffusion coefficients divided by the square root of the number of blocks.
    } 
    \label{fig:figure_2}
\end{figure*}

To reduce computational expense, both the simulation time and supercell size were systematically converged (see Figure \ref{fig:figure_2}); from these tests, a 672-atom supercell and a simulation length of 500~ps simulations were deemed sufficient.

\subsubsection{Reference values for Task 3 ($\sigma_{\text{RT}}$ of \ce{Li7P3S11})}
A comprehensive summary of reported experimental and computational values for the room-temperature ionic conductivity ($\sigma_{\text{RT}}$) of \ce{Li7P3S11} is provided in Table \ref{tab:task3_ref}. Experimental data were compiled from the review by Kudu et al.~\cite{kudu_review_2018}, where a detailed description of experimental synthesis conditions can be found, while computational references were collected independently in this work. 

Experimental investigations on \ce{Li7P3S11} have employed a variety of synthesis routes, including solid-state reactions, mechanochemical (ball-milling) methods, and wet-chemical techniques\cite{kudu_review_2018}. However, direct comparison between experimental and computational values of $\sigma_{\text{RT}}$ remains challenging for several reasons. Fully crystalline \ce{Li7P3S11} is challenging to synthesize experimentally, meaning that samples usually contain a significant proportion of amorphous phase. The mixture of phases, and the resulting introduction of grain boundaries has been shown to strongly influence the ionic conductivity \cite{minami_crystallization_2011, seino_analysis_2015}, and is largely responsible for the wide range of reported experimental conductivities. 

In contrast, computational studies typically simulate the intrinsic bulk conductivity of defect-free, fully crystalline \ce{Li7P3S11} under periodic boundary conditions. Since these models inherently neglect grain boundary effects and interfaces, direct comparison of experimental and computational values is of limited value. Additional sources of uncertainty in computational conductivity estimates arise from finite-size effects, limited simulation times, and the frequent use of the Nernst--Einstein relation to estimate $\sigma_{\text{RT}}$ from diffusivity (see Section S2.3.1). Moreover, due to the high cost of AIMD, simulations are often performed with much lower convergence criteria than those typically used for DFT labels for training or fine-tuning, such as reduced plane-wave cutoffs, coarser $k$-point grids, and less strict electronic convergence thresholds. These compromises can introduce numerical noise or systematic errors in forces and energies, which may affect diffusion behavior and lead to deviations in predicted conductivities. Together, these factors contribute to systematic differences between experimental and theoretical $\sigma_{\text{RT}}$ values; as such, quantitative agreement between experimental and computationally-derived conductivity data should not be expected.

For these reasons, we do not make direct comparisons between our MLIP-derived conductivities and experimental values. Instead, we consider predicted intrinsic bulk conductivities in the range of experiment to AIMD values to be reasonable, and as such, draw attention to ref.~\citenum{seino_sulphide_2014}, reporting the highest known experimental conductivity, and ref.~\citenum{chu_insights_2016}, reporting a representative AIMD value computed with the same exchange--correlation functional (PBEsol) as used in the present work.

\begin{table}[h]
    \caption{Room-temperature ionic conductivities ($\sigma_{\text{RT}}$) and activation energies ($E_{\text{a}}$) for \ce{Li7P3S11} reported in the literature. For experimental studies, the synthesis method and the phase type obtained (glass, glass-ceramic, or crystalline) is indicated (collected from ref.~\citenum{kudu_review_2018}). For computational studies, the simulation method and the phase modeled is noted.}
    \centering
    \begin{tabular}{ccccc}
        \hline
        \textbf{Ref.} & \textbf{Method} & \textbf{Phase} & \textbf{$\sigma_{\text{RT}}$ (mS/cm)} & \textbf{$E_{\text{a}}$ (eV)} \\ \hline
        \citenum{seino_sulphide_2014} & Solid-state & Glass-ceramic & 0.08   & --   \\ 
        \citenum{seino_sulphide_2014} & Solid-state & Glass-ceramic & 1.4   & 0.50  \\ 
        \citenum{seino_sulphide_2014} & Solid-state & Glass-ceramic  & 17   & 0.17  \\ 
        \citenum{chu_insights_2016} & Solid-state  & Glass-ceramic & 1.3   & 0.21 \\ 
        \citenum{chu_insights_2016} & Solid-state  & Glass-ceramic & 12   & 0.18 \\ \hline
        \citenum{hayashi_preparation_2004} & Mechanochemical & Glass  & 0.04   & 0.41 \\ 
        \citenum{dietrich_lithium_2017} & Mechanochemical & Glass   & 0.037   & 0.45 \\ 
        \citenum{mizuno_high_2006} & Mechanochemical & Glass-ceramic  & 3.2   & 0.12 \\ 
        \citenum{wenzel_interphase_2016} & Mechanochemical  & Glass & 0.05   & 0.38 \\
        \citenum{wenzel_interphase_2016} & Mechanochemical & Crystal  & 4   & 0.29 \\
        \citenum{busche_situ_2016} & Mechanochemical & Glass  & 0.081   & 0.43 \\
        \citenum{busche_situ_2016} & Mechanochemical  & Crystal & 8.6   & 0.29 \\ \hline
        \citenum{ito_synthesis_2014} & Wet chemistry & Glass-ceramic & 0.27   & 0.39  \\ 
        \citenum{wang_mechanism_2018} & Wet chemistry & Glass-ceramic & 0.87   & 0.37 \\ 
        \citenum{calpa_preparation_2018} & Wet chemistry & Glass-ceramic & 0.011   & -- \\ 
        \citenum{calpa_preparation_2018} & Wet chemistry & Glass-ceramic & 1.0   & 0.13 \\ \hline
        \hline
        \citenum{chu_insights_2016} & AIMD (PBE) & Crystal & 57.0  & 0.19 \\ 
        \citenum{chu_insights_2016} & AIMD (PBEsol) & Crystal & 61.0  & -- \\ 
        \citenum{baba_structure_2016} & AIMD (PBEsol) & Glass & 0.082  & -- \\ 
        \citenum{wang_computational_2017} & AIMD (PBE) & Crystal & 45.7 & 0.19 \\
        \citenum{chang_super-ionic_2018} & AIMD (PBE) & Crystal & 72.0  & 0.17 \\ 
        \citenum{sadowski_computational_2020} & AIMD (PBE) & Crystal & 84.0  & 0.17 \\
        \citenum{ohkubo_conduction_2020} & AIMD & Glass & 1.8  & -- \\ 
        \citenum{ohkubo_conduction_2020} & AIMD & Crystal & 240.0  & -- \\ 
        \hline
    \end{tabular}
    \label{tab:task3_ref}
\end{table}

\subsection{Task 4: Robustness}
A series of 100~ps NPT annealing runs was carried out for each MLIP model, across a $7 \times 7$ grid of $(T, P)$ conditions spanning temperatures of 1000, 2000, 4000, 8000, 16\,000, 32\,000, and 64\,000~K, and pressures of $10^6$, $10^7$, $10^8$, $10^9$, $10^{10}$, $10^{11}$, and $10^{12}$~Pa. Simulations used a 1~fs timestep, \(t_{\text{damp}}^{(T)} = 100\)~fs, and \(t_{\text{damp}}^{(p)} = 1000\)~fs. The starting point was an approximately cubic ($a \approx b \approx c$) 1008-atom random-hard-sphere \ce{Li7P3S11} structure generated with the \texttt{buildcell} code \cite{pickard_ab_2011}, pre-relaxed in a fixed cell with the corresponding potential using the BFGS optimizer in ASE\cite{hjorth_larsen_atomic_2017} until $|f_\text{max}| < 0.05$~eV/Å.
Analysis scripts are provided for both success criteria: (i) simulation survival and (ii) the number of close-contact events.

\clearpage

\section{Experiments}
\subsection{Benchmarking Graph-Based MLIPs}
Benchmark results for Tasks 1 and 2 were obtained by averaging predictions from five models trained with different random seeds (see Section S3.4.4.). For Task 3, a single representative MACE model from these five was selected, and three sets of annealing runs were performed with different random seeds for initializing atomic velocities. The resulting $\sigma_{298}$ predictions were then averaged across these three repeats.

\subsection{Fine-tuning Foundational Models with LiPS-25}
\begin{figure*}[h]
    \centering
    \includegraphics[width=\linewidth]{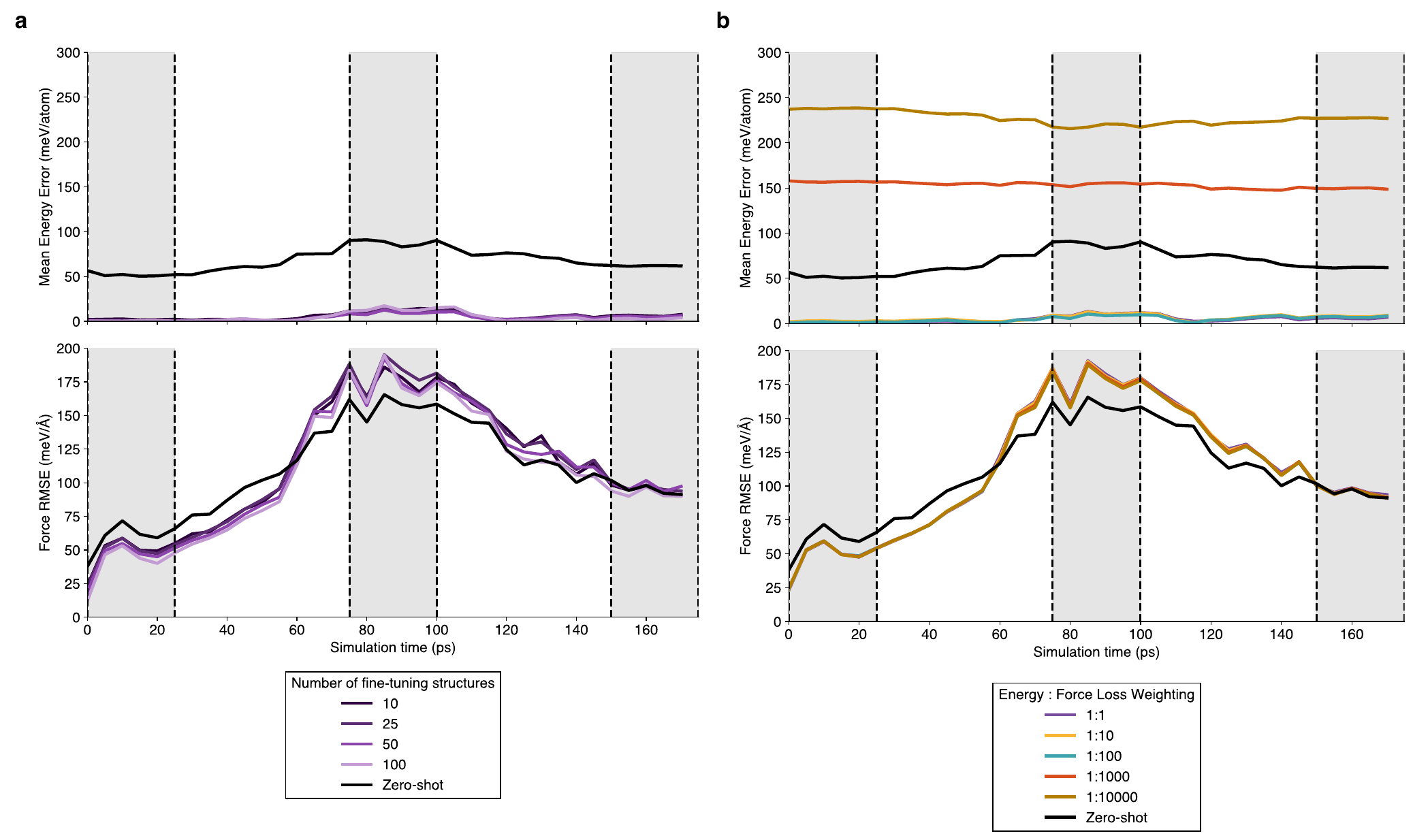}
    \caption{Hyperparameter optimization for the fine-tuning protocol, shown for MACE-OMAT-0 as a representative foundation model. Results are averaged over five repeats. (a) Effect of fine-tuning dataset size on energy and force errors: 25 \ce{Li7P3S11} structures are sufficient, with little improvement from larger datasets. Tests were conducted using a 1:1 energy:force loss ratio. (b) Effect of varying the energy:force loss weighting: increasing the force contribution yields negligible improvements in force accuracy but substantially worsens energy predictions. Fine-tuning was performed on 25 structures.}
    \label{fig:figure_3}
\end{figure*}

To determine the optimal hyperparameters for the fine-tuning protocol described in Section S3.4.4, we varied the learning rate, the relative weighting of energy and force terms in the loss function, and the number of \ce{Li7P3S11} structures in the fine-tuning dataset. Learning rates of 0.01, 0.001, and 0.0001 were tested: a setting of 0.01 led to significantly worse force predictions (particularly in the liquid regime), while 0.001 and 0.0001 performed comparably, with 0.0001 selected as optimal. Figure \ref{fig:figure_3}a shows the effect of dataset size, indicating that fine-tuning on 25 structures is sufficient, with little to no improvement from larger datasets. Figure \ref{fig:figure_3}b shows the effect of varying the energy:force weighting: increasing the force contribution gave negligible gains in force accuracy but substantially degraded energy predictions. Accordingly, a 1:1 weighting was adopted.

Each foundation model was fine-tuned with five different random seeds, and energy and force predictions from the resulting five models were averaged to produce the values shown in Figure 5.

\subsection{Ionic Conductivities from Fine-Tuned Models}
Ionic conductivities were computed following the protocol described in Section S2.3.2. For the fine-tuned models shown in Figure 5 (trained on 25 \ce{Li7P3S11} structures), one training repeat was selected to perform three independent annealing runs, each with a different random seed for initializing atomic velocities, and the resulting $\sigma_{298}$ values were averaged. The same procedure was applied to the zero-shot models, averaging the results from three anneals with distinct velocity seeds.

\subsection{Computational Details}
\subsubsection{DFT computations}
For the construction of the LiPS-25 dataset, DFT reference computations were performed using VASP 6.4.3\cite{kresse_efficient_1996, kresse_efficiency_1996, kresse_ab_1994, kresse_ab_1995} with the PBEsol exchange--correlation functional \cite{perdew_restoring_2008} and projector augmented-wave pseudopotentials  (\texttt{PAW\_PBE Li\_sv 10Sep2004}, \texttt{PAW\_PBE P 06Sep2000}, and \texttt{PAW\_PBE S 06Sep2000}) \cite{blochl_projector_1994, kresse_ultrasoft_1999}. A plane-wave cutoff of 1000 eV and an energy tolerance of $10^{-8}$ eV per cell were chosen for SCF convergence. Brillouin-zone sampling was carried out using automatically generated $k$-point grids with a maximum spacing of 0.2~\AA{}$^{-1}$.

For the initial structural optimization of crystalline structures in Iter0, the same plane-wave cutoff energy (1000 eV) was used, with an SCF energy tolerance of $10^{-6}$ eV per cell. The convergence criterion for ionic relaxation was a force tolerance of $10^{-2}$ eV~\AA{}$^{-1}$, and reciprocal space was sampled using automatically generated $k$-point grids with a spacing of 0.2~\AA{}$^{-1}$.

Ab initio molecular dynamics (AIMD) simulations were performed using VASP 6.3.2 to generate structures for the Iter0 dataset. These calculations employed a plane-wave cutoff of 400 eV, an SCF energy tolerance of $10^{-5}$ eV per cell, and $\Gamma$-point sampling only. The simulations were carried out in the NVT ensemble using a Nosé–Hoover thermostat, with a timestep of 1 fs.

\subsubsection{NequIP fitting}
For dataset augmentation beyond Iter0, we employed NequIP \cite{batzner_e3-equivariant_2022}. All models were trained on an NVIDIA RTX A6000 GPU in \texttt{float32} precision. The training hyperparameters were: cutoff radius $r_\text{max}=4.5$~\AA{}; $l_\text{max}=2$; 32 features (including both even and odd); and 6 interaction layers. The invariant radial networks operated on a trainable Bessel basis of size 8 and were implemented with two hidden layers of 64 neurons, using SiLU nonlinearities. 

Training used a learning rate of 0.001, a batch size of 50, and a loss function with equal weighting between energy and force terms. The learning rate was reduced by a factor of 0.5 if the validation loss did not improve for 100 epochs. Early stopping was applied if the validation loss failed to decrease by at least 0.005 over 40 epochs, otherwise the maximum number of epochs was set to 100,000. This model was then used to drive melt–quench simulations of the seven key compositions at each iteration.

At each iteration, a committee of five NequIP models with identical hyperparameters was trained on different random 50\% subsets of the available dataset. This ensemble was used to identify and select the most uncertain MD frames for inclusion in the next training round.

\subsubsection{MACE fitting}
For the hyperparameter sweep shown in Figure 3, five MACE models were trained with different random seeds for each set of hyperparameters. All models were trained on an NVIDIA A100 GPU using \texttt{float32} precision. Except for the cutoff sweep, a cutoff of 6~\AA{} was used for all models along with the hyperparameters specified. Training was performed using the MACE implementation provided in \texttt{graph-pes} \cite{gardner_graph-pes_2024}.

Hyperparameters not included in the sweep were kept at their default values as defined in \texttt{graph-pes}. Training was performed with a learning rate of 0.001 and a batch size of 32, except for the largest models where a reduced batch size of 5 was used due to memory constraints. The loss function combined energy and force terms in a 1:1 ratio. The learning rate was decreased by a factor of 0.8 if the validation loss did not improve over 25 epochs. Models were trained for a maximum of 1000 epochs.

\subsubsection{Fine-tuning}
The foundation models used in this study were as follows: for the MACE family \cite{batatia_foundation_2025}, MACE-MP-0b3, MACE-MPA-0, MACE-OMAT-0, and MACE-MATPES-0 (available at \url{https://github.com/ACEsuit/mace}); for the Orb family \cite{neumann_orb_2024, rhodes_orb-v3_2025}, Orb-v2, Orb-v3-direct-inf-mpa, and Orb-v3-direct-inf-omat (available at \url{https://github.com/orbital-materials/orb-models}); and for the MatterSim family \cite{yang_mattersim_2024}, MatterSim-1m and MatterSim-5m (available at \url{https://github.com/microsoft/mattersim}). Fine-tuning of the foundation models was carried out using the \texttt{graph-pes} package \cite{gardner_graph-pes_2024} following a “naive” protocol, in which pre-trained weights are updated directly using the fine-tuning dataset. All models were trained on an NVIDIA RTX A6000 GPU in \texttt{float32} precision. A 6~\AA{} cutoff with a learnable offset was used. Training employed a learning rate of 0.0001 and a batch size equal to the number of structures used for fine-tuning (25). The same loss function and learning rate schedule as described for MACE fitting in Section S3.4.3 were applied.

\clearpage

\setstretch{1.0}

\providecommand{\latin}[1]{#1}
\makeatletter
\providecommand{\doi}
  {\begingroup\let\do\@makeother\dospecials
  \catcode`\{=1 \catcode`\}=2 \doi@aux}
\providecommand{\doi@aux}[1]{\endgroup\texttt{#1}}
\makeatother
\providecommand*\mcitethebibliography{\thebibliography}
\csname @ifundefined\endcsname{endmcitethebibliography}  {\let\endmcitethebibliography\endthebibliography}{}